\documentclass[11pt]{article}
\usepackage{latexsym,a4,epsfig}

\newlength{\dinwidth}
\newlength{\dinmargin}
\def\ga{\mathrel{\raise.3ex\hbox{$>$\kern-.75em\lower1ex\hbox{$\sim$}}}}
\def\la{\mathrel{\raise.3ex\hbox{$<$\kern-.75em\lower1ex\hbox{$\sim$}}}}
\def\gev{{\rm \, Ge\kern-0.125em V}}
\def\tev{{\rm \, Te\kern-0.125em V}}
\newcommand{\gevc}{\ensuremath{\mathrm{GeV}/c}}

\newcommand{\rts}{\sqrt{s}}

\def \ggqq {\gamma\gamma \rightarrow q\overline{q}}

\newcommand{\Mcha}{M_{\chi^\pm}}
\newcommand{\Mchi}{M_\chi}

\newcommand{\ZZg} {\mathrm{ZZ^{*}/\gamma^{*}}}

\newcommand {\susy}{supersymmetry}
\newcommand {\susyk}{supersymmetric}
\newcommand{\gaga}{\ensuremath{\gamma\gamma}}
\newcommand{\gagahh}{\ensuremath{\gamma\gamma\rightarrow hadrons }}
\newcommand {\qqg} {{\mathrm q\overline{\mathrm q}}\gamma}
\newcommand{\tanb}{\ensuremath{\tan\beta}}
\newcommand{\dm}{\ensuremath{\Delta M}}
\newcommand{\pcecm}{\ensuremath{\%\sqrt s}}
\newcommand{\whad}{\ensuremath{W_{\mathrm{had}}}}
\newcommand{\evis}{\ensuremath{E_{\mathrm{vis}}}}
\newcommand{\mvis}{\ensuremath{M_{\mathrm{vis}}}}
\newcommand{\nch}{\ensuremath{N_{\mathrm{ch}}}}
\newcommand{\Nbarnf}{\ensuremath{\bar N_{95}}}
\newcommand{\pt}{{\ensuremath {P_{T}}}}
\newcommand{\ptlep}{{\ensuremath {P_{T}^{lepton}}}}
\newcommand{\elow}{{\ensuremath {E_{12}}}}
\newcommand{\thd}{{\ensuremath {\theta_{2}}}}
\newcommand{\thu}{{\ensuremath {\theta_{1}}}}
\newcommand{\fu}{{\ensuremath {f_{1}}}}
\newcommand{\fd}{{\ensuremath {f_{2}}}}

\newcommand{\fnh}{{\ensuremath {F_{NH}}}}
\newcommand{\elepu}{{\ensuremath {E_{\ell 1}}}}
\newcommand{\elepd}{{\ensuremath {E_{\ell 2}}}}
\newcommand{\egam}{{\ensuremath {E_{\gamma}}}}
\newcommand{\ewref}{{\ensuremath {E_{\mathrm w}^{\mathrm{ref}}}}}
\newcommand{\efwd}{{\ensuremath {E_{30^\circ}}}}
\newcommand{\elref}{{\ensuremath {E_{\ell}^{\mathrm{ref}}}}}
\newcommand{\invb}{{\ensuremath {Inv}\mathcal{B}}}
\newcommand{\mmw}{{\ensuremath {M}_{W}}}
\newcommand{\adt}{{\ensuremath {\alpha}_{23}}}
\newcommand{\aref}{{\ensuremath {\alpha_{\mathrm{ref}}}}}
\newcommand{\aco}{{\ensuremath {\Delta\phi}}}
\newcommand{\acop}{{\ensuremath {\Delta\phi}}}
\newcommand{\act}{{\ensuremath {\Delta\phi_{\mathrm T}}}}
\newcommand{\acopt}{\ensuremath{\Delta\phi_{\mathrm T}}}
\newcommand{\wmiss}{\ensuremath{M_{\mathrm{miss}}}}
\newcommand{\mmiss}{\ensuremath {M_{\mathrm{miss}}}}

\newcommand{\phmiss}{\ensuremath{\phi_{\mathrm{miss}}}}
\newcommand{\tsc}{\ensuremath {\theta_{\mathrm{scat}}}}
\newcommand{\thscat}{\ensuremath{\theta_{\mathrm{scat}}}}
\newcommand{\thpoint}{\ensuremath{\theta_{\mathrm{point}}}}
\newcommand{\tpt}{\ensuremath {\theta_{\mathrm{point}}}}
\newcommand{\ctmis}{\ensuremath {\cos\theta_{\mathrm{miss}}}}
\newcommand{\cmiss}{\ensuremath{|\cos\theta_{\mathrm{miss}}|}}
\newcommand{\ptnonh}{\ensuremath{P_T^{\overline{NH}}}}
\newcommand{\ptnh}{{\ensuremath {P_{T}^{\overline{NH}}}}}
\newcommand{\ewedge}{\ensuremath{E_{\mathrm w}^{30}}}
\newcommand{\ewed}{{\ensuremath {E_{\mathrm w}^{30}}}}
\newcommand{\elept}{\ensuremath{E_{\ell}}}
\newcommand{\elep}{{\ensuremath {E_{\ell}}}}
\newcommand{\elcone}{\ensuremath{E_{\ell}^{30}}}
\newcommand{\ewedl}{{\ensuremath {E_{\ell}^{30}}}}

\newcommand{\M}{\ensuremath M_{2}}
\newcommand{\Mp}{\ensuremath M_{2}}
\newcommand{\sigbg}{\ensuremath \sigma_{\mathrm{bg}}}
\newcommand{\ww}   {\ensuremath\mathrm {WW}}
\newcommand{\zz}   {\ensuremath\mathrm Z\gamma^{*}}
\newcommand{\ewnu} {\ensuremath\mathrm{We}\nu}

\newcommand{\gagall}{\ensuremath{\gamma\gamma\rightarrow \ell\ell }}
\newcommand{\Pstaup}{\ensuremath{\widetilde{\tau}_{1}}}
\newcommand{\mzero}{\ensuremath  m_{0}}
\newcommand{\msnu}{\ensuremath  M_{\tilde{\nu}}}
\newcommand{\mcha}{\ensuremath  M_{\chi^{\pm}}}
\newcommand{\mchi}{\ensuremath  M_{\chi}}
\newcommand{\mchip}{\ensuremath  M_{\chi'}}
\newcommand{\atau}{\ensuremath  A_{\tau}}
\newcommand{\chsnu}{\ensuremath \PCha \rightarrow \ell \tilde{\nu}}
\newcommand{\chstau}{\ensuremath \PCha \rightarrow \tilde{\tau}_1 \nu}
\newcommand{\chlep}{\ensuremath \PCha \rightarrow \ell\nu\chi}
\newcommand{\chtau}{\ensuremath \PCha \rightarrow \tau\nu\chi}

\newcommand{\Tcsq}{\ensuremath{\mathrm{TeV}/c^2}}
\newcommand{\dedx}{\ensuremath{\mathrm{d}E/\mathrm{d}x}}

\newcommand{\PCha}{\ensuremath{\chi^{\pm}}}

\newcommand{\PSnu}{\ensuremath{\widetilde{\nu}}}
\newcommand{\PaSnu}{\ensuremath{\overline{\nu}}}
\newcommand{\Pnu}{\ensuremath{\nu}}
\newcommand{\Panu}{\ensuremath{\overline{\nu}}}

\newcommand{\Pchi}{\ensuremath{\chi}}
\newcommand{\Pchip}{\ensuremath{\chi'}}
\newcommand{\Pchipp}{\ensuremath{\chi''}}
\newcommand{\Pchippp}{\ensuremath{\chi'''}}
\newcommand{\Gcsq}{\ensuremath{\mathrm{GeV}/c^2}}
\newcommand{\tb}{\ensuremath{\tan\beta}}

\newcommand{\pbinv}{\ensuremath{\mathrm{pb}^{-1}}}
\newcommand{\epem}{\ensuremath{\mathrm{e^+e^-}}}

\begin{document}
\thispagestyle{empty}
~\\
\begin{center}
{\Large EUROPEAN LABORATORY FOR PARTICLE PHYSICS (CERN)}
\end{center}
\bigskip
\begin{flushright}
\date{} 
CERN-PPE/97-128 \\
  September 22, 1997 \\
\end{flushright}

\vskip 1.5cm
\begin{center}
{\Huge Searches for Charginos and Neutralinos\\}
\vskip 0.5cm
{\Huge in $\mathrm {e^{+}e^{-}}$ Collisions\\}
\vskip 0.5cm
{\Huge at $\rts$ = 161 and 172~GeV }
~~\\
\vspace{1cm}

{\Large The ALEPH Collaboration}
~~\\
\vspace{1.2cm}
\begin{quotation}
\begin{center}
{\bf Abstract}\\
\end{center}

\noindent

{\small
The data recorded by the ALEPH detector at centre-of-mass energies of 
161, 170, and 172~GeV are analysed for signals of chargino 
and neutralino production. 
No evidence of a signal is found, although candidate 
events consistent with the 
expectations from Standard Model processes are observed.
Limits at $95\%$ C.L. on the production
cross sections are derived and bounds on the parameters of
the Minimal 
Supersymmetric Standard Model are set. 
The lower limit on the mass of the lightest chargino 
is 85.5~$\Gcsq$ for gaugino-like charginos ($\mu = -500~\Gcsq$), and 
85.0~$\Gcsq$ for 
Higgsino-like  charginos ($\Mp = 500~\Gcsq$), for heavy sneutrinos 
($\msnu \geq 200~\Gcsq$) and $\tanb = \sqrt{2}$.
The effect of light sleptons on chargino and neutralino limits 
is investigated.
The assumptions of a universal slepton mass and a universal gaugino 
mass are relaxed, allowing
less model-dependent limits to be obtained.
 }

\end{quotation}

\vskip 0.5cm
\vskip 0.5cm
\small
\it {(submitted to Zeitschrift f\"ur Physik)}
\vskip 0.5cm
\end{center}


%
\vfill
\normalsize
\newpage
\pagestyle{plain}
\setcounter{page}{1}
\pretolerance=1000

\pagestyle{empty}
\newpage
\small
%
%
\newlength{\saveparskip}
\newlength{\savetextheight}
\newlength{\savetopmargin}
\newlength{\savetextwidth}
\newlength{\saveoddsidemargin}
\newlength{\savetopsep}
\setlength{\saveparskip}{\parskip}
\setlength{\savetextheight}{\textheight}
\setlength{\savetopmargin}{\topmargin}
\setlength{\savetextwidth}{\textwidth}
\setlength{\saveoddsidemargin}{\oddsidemargin}
\setlength{\savetopsep}{\topsep}
%
%
\setlength{\parskip}{0.0cm}
\setlength{\textheight}{25.0cm}
\setlength{\topmargin}{-1.5cm}
\setlength{\textwidth}{16 cm}
\setlength{\oddsidemargin}{-0.0cm}
\setlength{\topsep}{1mm}
\pretolerance=10000
\centerline{\large\bf The ALEPH Collaboration}
\footnotesize
\vspace{0.5cm}
{\raggedbottom
\begin{sloppypar}
\samepage\noindent
R.~Barate,
D.~Buskulic,
D.~Decamp,
P.~Ghez,
C.~Goy,
J.-P.~Lees,
A.~Lucotte,
M.-N.~Minard,
J.-Y.~Nief,
B.~Pietrzyk
\nopagebreak
\begin{center}
\parbox{15.5cm}{\sl\samepage
Laboratoire de Physique des Particules (LAPP), IN$^{2}$P$^{3}$-CNRS,
74019 Annecy-le-Vieux Cedex, France}
\end{center}\end{sloppypar}
\vspace{2mm}
\begin{sloppypar}
\noindent
M.P.~Casado,
M.~Chmeissani,
P.~Comas,
J.M.~Crespo,
M.~Delfino, 
E.~Fernandez,
M.~Fernandez-Bosman,
Ll.~Garrido,$^{15}$
A.~Juste,
M.~Martinez,
G.~Merino,
R.~Miquel,
Ll.M.~Mir,
C.~Padilla,
I.C.~Park,
A.~Pascual,
J.A.~Perlas,
I.~Riu,
F.~Sanchez
\nopagebreak
\begin{center}
\parbox{15.5cm}{\sl\samepage
Institut de F\'{i}sica d'Altes Energies, Universitat Aut\`{o}noma
de Barcelona, 08193 Bellaterra (Barcelona), Spain$^{7}$}
\end{center}\end{sloppypar}
\vspace{2mm}
\begin{sloppypar}
\noindent
A.~Colaleo,
D.~Creanza,
M.~de~Palma,
G.~Gelao,
G.~Iaselli,
G.~Maggi,
M.~Maggi,
N.~Marinelli,
S.~Nuzzo,
A.~Ranieri,
G.~Raso,
F.~Ruggieri,
G.~Selvaggi,
L.~Silvestris,
P.~Tempesta,
A.~Tricomi,$^{3}$
G.~Zito
\nopagebreak
\begin{center}
\parbox{15.5cm}{\sl\samepage
Dipartimento di Fisica, INFN Sezione di Bari, 70126
Bari, Italy}
\end{center}\end{sloppypar}
\vspace{2mm}
\begin{sloppypar}
\noindent
X.~Huang,
J.~Lin,
Q. Ouyang,
T.~Wang,
Y.~Xie,
R.~Xu,
S.~Xue,
J.~Zhang,
L.~Zhang,
W.~Zhao
\nopagebreak
\begin{center}
\parbox{15.5cm}{\sl\samepage
Institute of High-Energy Physics, Academia Sinica, Beijing, The People's
Republic of China$^{8}$}
\end{center}\end{sloppypar}
\vspace{2mm}
\begin{sloppypar}
\noindent
D.~Abbaneo,
R.~Alemany,
A.O.~Bazarko,$^{1}$
U.~Becker,
P.~Bright-Thomas,
M.~Cattaneo,
F.~Cerutti,
G.~Dissertori,
H.~Drevermann,
R.W.~Forty,
M.~Frank,
F.~Gianotti,
R.~Hagelberg,
J.B.~Hansen,
J.~Harvey,
P.~Janot,
B.~Jost,
E.~Kneringer,
I.~Lehraus,
P.~Mato,
A.~Minten,
L.~Moneta,
A.~Pacheco,
J.-F.~Pusztaszeri,$^{20}$
F.~Ranjard,
G.~Rizzo,
L.~Rolandi,
D.~Rousseau,
D.~Schlatter,
M.~Schmitt,
O.~Schneider,
W.~Tejessy,
F.~Teubert,
I.R.~Tomalin,
H.~Wachsmuth,
A.~Wagner$^{21}$
\nopagebreak
\begin{center}
\parbox{15.5cm}{\sl\samepage
European Laboratory for Particle Physics (CERN), 1211 Geneva 23,
Switzerland}
\end{center}\end{sloppypar}
\vspace{2mm}
\begin{sloppypar}
\noindent
Z.~Ajaltouni,
A.~Barr\`{e}s,
C.~Boyer,
A.~Falvard,
C.~Ferdi,
P.~Gay,
C~.~Guicheney,
P.~Henrard,
J.~Jousset,
B.~Michel,
S.~Monteil,
J-C.~Montret,
D.~Pallin,
P.~Perret,
F.~Podlyski,
J.~Proriol,
P.~Rosnet,
J.-M.~Rossignol
\nopagebreak
\begin{center}
\parbox{15.5cm}{\sl\samepage
Laboratoire de Physique Corpusculaire, Universit\'e Blaise Pascal,
IN$^{2}$P$^{3}$-CNRS, Clermont-Ferrand, 63177 Aubi\`{e}re, France}
\end{center}\end{sloppypar}
\vspace{2mm}
\begin{sloppypar}
\noindent
T.~Fearnley,
J.D.~Hansen,
J.R.~Hansen,
P.H.~Hansen,
B.S.~Nilsson,
B.~Rensch,
A.~W\"a\"an\"anen
\begin{center}
\parbox{15.5cm}{\sl\samepage
Niels Bohr Institute, 2100 Copenhagen, Denmark$^{9}$}
\end{center}\end{sloppypar}
\vspace{2mm}
\begin{sloppypar}
\noindent
G.~Daskalakis,
A.~Kyriakis,
C.~Markou,
E.~Simopoulou,
A.~Vayaki
\nopagebreak
\begin{center}
\parbox{15.5cm}{\sl\samepage
Nuclear Research Center Demokritos (NRCD), Athens, Greece}
\end{center}\end{sloppypar}
\vspace{2mm}
\begin{sloppypar}
\noindent
A.~Blondel,
J.C.~Brient,
F.~Machefert,
A.~Roug\'{e},
M.~Rumpf,
A.~Valassi,$^{6}$
H.~Videau
\nopagebreak
\begin{center}
\parbox{15.5cm}{\sl\samepage
Laboratoire de Physique Nucl\'eaire et des Hautes Energies, Ecole
Polytechnique, IN$^{2}$P$^{3}$-CNRS, 91128 Palaiseau Cedex, France}
\end{center}\end{sloppypar}
\vspace{2mm}
\begin{sloppypar}
\noindent
T.~Boccali,
E.~Focardi,
G.~Parrini,
K.~Zachariadou
\nopagebreak
\begin{center}
\parbox{15.5cm}{\sl\samepage
Dipartimento di Fisica, Universit\`a di Firenze, INFN Sezione di Firenze,
50125 Firenze, Italy}
\end{center}\end{sloppypar}
\vspace{2mm}
\begin{sloppypar}
\noindent
R.~Cavanaugh,
M.~Corden,
C.~Georgiopoulos,
T.~Huehn,
D.E.~Jaffe
\nopagebreak
\begin{center}
\parbox{15.5cm}{\sl\samepage
Supercomputer Computations Research Institute,
Florida State University,
Tallahassee, FL 32306-4052, USA $^{13,14}$}
\end{center}\end{sloppypar}
\vspace{2mm}
\begin{sloppypar}
\noindent
A.~Antonelli,
G.~Bencivenni,
G.~Bologna,$^{4}$
F.~Bossi,
P.~Campana,
G.~Capon,
D.~Casper,
V.~Chiarella,
G.~Felici,
P.~Laurelli,
G.~Mannocchi,$^{5}$
F.~Murtas,
G.P.~Murtas,
L.~Passalacqua,
M.~Pepe-Altarelli
\nopagebreak
\begin{center}
\parbox{15.5cm}{\sl\samepage
Laboratori Nazionali dell'INFN (LNF-INFN), 00044 Frascati, Italy}
\end{center}\end{sloppypar}
\vspace{2mm}
\begin{sloppypar}
\noindent
L.~Curtis,
S.J.~Dorris,
A.W.~Halley,
I.G.~Knowles,
J.G.~Lynch,
V.~O'Shea,
C.~Raine,
J.M.~Scarr,
K.~Smith,
P.~Teixeira-Dias,
A.S.~Thompson,
E.~Thomson,
F.~Thomson,
R.M.~Turnbull
\nopagebreak
\begin{center}
\parbox{15.5cm}{\sl\samepage
Department of Physics and Astronomy, University of Glasgow, Glasgow G12
8QQ,United Kingdom$^{10}$}
\end{center}\end{sloppypar}
\vspace{2mm}
\begin{sloppypar}
\noindent
O.~Buchm\"uller,
S.~Dhamotharan,
C.~Geweniger,
G.~Graefe,
P.~Hanke,
G.~Hansper,
V.~Hepp,
E.E.~Kluge,
A.~Putzer,
J.~Sommer,
K.~Tittel,
S.~Werner,
M.~Wunsch
\begin{center}
\parbox{15.5cm}{\sl\samepage
Institut f\"ur Hochenergiephysik, Universit\"at Heidelberg, 69120
Heidelberg, Fed.\ Rep.\ of Germany$^{16}$}
\end{center}\end{sloppypar}
\vspace{2mm}
\begin{sloppypar}
\noindent
R.~Beuselinck,
D.M.~Binnie,
W.~Cameron,
P.J.~Dornan,
M.~Girone,
S.~Goodsir,
E.B.~Martin,
P.~Morawitz,
A.~Moutoussi,
J.~Nash,
J.K.~Sedgbeer,
P.~Spagnolo,
A.M.~Stacey,
M.D.~Williams
\nopagebreak
\begin{center}
\parbox{15.5cm}{\sl\samepage
Department of Physics, Imperial College, London SW7 2BZ,
United Kingdom$^{10}$}
\end{center}\end{sloppypar}
\vspace{2mm}
\begin{sloppypar}
\noindent
V.M.~Ghete,
P.~Girtler,
D.~Kuhn,
G.~Rudolph
\nopagebreak
\begin{center}
\parbox{15.5cm}{\sl\samepage
Institut f\"ur Experimentalphysik, Universit\"at Innsbruck, 6020
Innsbruck, Austria$^{18}$}
\end{center}\end{sloppypar}
\vspace{2mm}
\begin{sloppypar}
\noindent
A.P.~Betteridge,
C.K.~Bowdery,
P.G.~Buck,
P.~Colrain,
G.~Crawford,
A.J.~Finch,
F.~Foster,
G.~Hughes,
R.W.L.~Jones,
T.~Sloan,
E.P.~Whelan,
M.I.~Williams
\nopagebreak
\begin{center}
\parbox{15.5cm}{\sl\samepage
Department of Physics, University of Lancaster, Lancaster LA1 4YB,
United Kingdom$^{10}$}
\end{center}\end{sloppypar}
\vspace{2mm}
\begin{sloppypar}
\noindent
I.~Giehl,
C.~Hoffmann,
K.~Jakobs,
K.~Kleinknecht,
G.~Quast,
B.~Renk,
E.~Rohne,
H.-G.~Sander,
P.~van~Gemmeren,
C.~Zeitnitz
\nopagebreak
\begin{center}
\parbox{15.5cm}{\sl\samepage
Institut f\"ur Physik, Universit\"at Mainz, 55099 Mainz, Fed.\ Rep.\
of Germany$^{16}$}
\end{center}\end{sloppypar}
\vspace{2mm}
\begin{sloppypar}
\noindent
J.J.~Aubert,
C.~Benchouk,
A.~Bonissent,
G.~Bujosa,
J.~Carr,
P.~Coyle,
C.~Diaconu,
A.~Ealet,
D.~Fouchez,
N.~Konstantinidis,
O.~Leroy,
F.~Motsch,
P.~Payre,
M.~Talby,
A.~Sadouki,
M.~Thulasidas,
A.~Tilquin,
K.~Trabelsi
\nopagebreak
\begin{center}
\parbox{15.5cm}{\sl\samepage
Centre de Physique des Particules, Facult\'e des Sciences de Luminy,
IN$^{2}$P$^{3}$-CNRS, 13288 Marseille, France}
\end{center}\end{sloppypar}
\vspace{2mm}
\begin{sloppypar}
\noindent
M.~Aleppo, 
M.~Antonelli,
F.~Ragusa
\nopagebreak
\begin{center}
\parbox{15.5cm}{\sl\samepage
Dipartimento di Fisica, Universit\`a di Milano e INFN Sezione di
Milano, 20133 Milano, Italy.}
\end{center}\end{sloppypar}
\vspace{2mm}
\begin{sloppypar}
\noindent
R.~Berlich,
W.~Blum,
V.~B\"uscher,
H.~Dietl,
G.~Ganis,
C.~Gotzhein,
H.~Kroha,
G.~L\"utjens,
G.~Lutz,
W.~M\"anner,
H.-G.~Moser,
R.~Richter,
A.~Rosado-Schlosser,
S.~Schael,
R.~Settles,
H.~Seywerd,
R.~St.~Denis,
H.~Stenzel,
W.~Wiedenmann,
G.~Wolf
\nopagebreak
\begin{center}
\parbox{15.5cm}{\sl\samepage
Max-Planck-Institut f\"ur Physik, Werner-Heisenberg-Institut,
80805 M\"unchen, Fed.\ Rep.\ of Germany\footnotemark[16]}
\end{center}\end{sloppypar}
\vspace{2mm}
\begin{sloppypar}
\noindent
J.~Boucrot,
O.~Callot,$^{12}$
S.~Chen,
M.~Davier,
L.~Duflot,
J.-F.~Grivaz,
Ph.~Heusse,
A.~H\"ocker,
A.~Jacholkowska,
D.W.~Kim,$^{2}$
F.~Le~Diberder,
J.~Lefran\c{c}ois,
A.-M.~Lutz,
M.~Marumi,
M.-H.~Schune,
L.~Serin,
E.~Tournefier,
J.-J.~Veillet,
I.~Videau,
D.~Zerwas
\nopagebreak
\begin{center}
\parbox{15.5cm}{\sl\samepage
Laboratoire de l'Acc\'el\'erateur Lin\'eaire, Universit\'e de Paris-Sud,
IN$^{2}$P$^{3}$-CNRS, 91405 Orsay Cedex, France}
\end{center}\end{sloppypar}
\vspace{2mm}
\begin{sloppypar}
\noindent
\samepage
P.~Azzurri,
G.~Bagliesi,$^{12}$
S.~Bettarini,
C.~Bozzi,
G.~Calderini,
V.~Ciulli,
R.~Dell'Orso,
R.~Fantechi,
I.~Ferrante,
A.~Giassi,
A.~Gregorio,
F.~Ligabue,
A.~Lusiani,
P.S.~Marrocchesi,
A.~Messineo,
F.~Palla,
G.~Sanguinetti,
A.~Sciab\`a,
G.~Sguazzoni,
J.~Steinberger,
R.~Tenchini,
C.~Vannini,
A.~Venturi,
P.G.~Verdini
\samepage
\begin{center}
\parbox{15.5cm}{\sl\samepage
Dipartimento di Fisica dell'Universit\`a, INFN Sezione di Pisa,
e Scuola Normale Superiore, 56010 Pisa, Italy}
\end{center}\end{sloppypar}
\vspace{2mm}
\begin{sloppypar}
\noindent
G.A.~Blair,
L.M.~Bryant,
J.T.~Chambers,
M.G.~Green,
T.~Medcalf,
P.~Perrodo,
J.A.~Strong,
J.H.~von~Wimmersperg-Toeller
\nopagebreak
\begin{center}
\parbox{15.5cm}{\sl\samepage
Department of Physics, Royal Holloway \& Bedford New College,
University of London, Surrey TW20 OEX, United Kingdom$^{10}$}
\end{center}\end{sloppypar}
\vspace{2mm}
\begin{sloppypar}
\noindent
D.R.~Botterill,
R.W.~Clifft,
T.R.~Edgecock,
S.~Haywood,
P.~Maley,
P.R.~Norton,
J.C.~Thompson,
A.E.~Wright
\nopagebreak
\begin{center}
\parbox{15.5cm}{\sl\samepage
Particle Physics Dept., Rutherford Appleton Laboratory,
Chilton, Didcot, Oxon OX11 OQX, United Kingdom$^{10}$}
\end{center}\end{sloppypar}
\vspace{2mm}
\begin{sloppypar}
\noindent
B.~Bloch-Devaux,
P.~Colas,
B.~Fabbro,
E.~Lan\c{c}on,
M.C.~Lemaire,
E.~Locci,
P.~Perez,
J.~Rander,
J.-F.~Renardy,
A.~Rosowsky,
A.~Roussarie,
J.~Schwindling,
A.~Trabelsi,
B.~Vallage
\nopagebreak
\begin{center}
\parbox{15.5cm}{\sl\samepage
CEA, DAPNIA/Service de Physique des Particules,
CE-Saclay, 91191 Gif-sur-Yvette Cedex, France$^{17}$}
\end{center}\end{sloppypar}
\vspace{2mm}
\begin{sloppypar}
\noindent
S.N.~Black,
J.H.~Dann,
H.Y.~Kim,
A.M.~Litke,
M.A. McNeil,
G.~Taylor
\nopagebreak
\begin{center}
\parbox{15.5cm}{\sl\samepage
Institute for Particle Physics, University of California at
Santa Cruz, Santa Cruz, CA 95064, USA$^{19}$}
\end{center}\end{sloppypar}
\pagebreak
\vspace{2mm}
\begin{sloppypar}
\noindent
C.N.~Booth,
C.A.J.~Brew,
S.~Cartwright,
F.~Combley,
M.S.~Kelly,
M.~Lehto,
J.~Reeve,
L.F.~Thompson
\nopagebreak
\begin{center}
\parbox{15.5cm}{\sl\samepage
Department of Physics, University of Sheffield, Sheffield S3 7RH,
United Kingdom$^{10}$}
\end{center}\end{sloppypar}
\vspace{2mm}
\begin{sloppypar}
\noindent
K.~Affholderbach,
A.~B\"ohrer,
S.~Brandt,
G.~Cowan,
J.~Foss,
C.~Grupen,
L.~Smolik,
F.~Stephan 
\nopagebreak
\begin{center}
\parbox{15.5cm}{\sl\samepage
Fachbereich Physik, Universit\"at Siegen, 57068 Siegen,
 Fed.\ Rep.\ of Germany$^{16}$}
\end{center}\end{sloppypar}
\vspace{2mm}
\begin{sloppypar}
\noindent
M.~Apollonio,
L.~Bosisio,
R.~Della~Marina,
G.~Giannini,
B.~Gobbo,
G.~Musolino
\nopagebreak
\begin{center}
\parbox{15.5cm}{\sl\samepage
Dipartimento di Fisica, Universit\`a di Trieste e INFN Sezione di Trieste,
34127 Trieste, Italy}
\end{center}\end{sloppypar}
\vspace{2mm}
\begin{sloppypar}
\noindent
J.~Putz,
J.~Rothberg,
S.~Wasserbaech,
R.W.~Williams
\nopagebreak
\begin{center}
\parbox{15.5cm}{\sl\samepage
Experimental Elementary Particle Physics, University of Washington, WA 98195
Seattle, U.S.A.}
\end{center}\end{sloppypar}
\vspace{2mm}
\begin{sloppypar}
\noindent
S.R.~Armstrong,
E.~Charles,
P.~Elmer,
D.P.S.~Ferguson,
Y.~Gao,
S.~Gonz\'{a}lez,
T.C.~Greening,
O.J.~Hayes,
H.~Hu,
S.~Jin,
P.A.~McNamara III,
J.M.~Nachtman,
J.~Nielsen,
W.~Orejudos,
Y.B.~Pan,
Y.~Saadi,
I.J.~Scott,
J.~Walsh,
Sau~Lan~Wu,
X.~Wu,
J.M.~Yamartino,
G.~Zobernig
\nopagebreak
\begin{center}
\parbox{15.5cm}{\sl\samepage
Department of Physics, University of Wisconsin, Madison, WI 53706,
USA$^{11}$}
\end{center}\end{sloppypar}
}
\footnotetext[1]{Now at Princeton University, Princeton, NJ 08544, U.S.A.}
\footnotetext[2]{Permanent address: Kangnung National University, Kangnung,
Korea.}
\footnotetext[3]{Also at Dipartimento di Fisica, INFN Sezione di Catania,
Catania, Italy.}
\footnotetext[4]{Also Istituto di Fisica Generale, Universit\`{a} di
Torino, Torino, Italy.}
\footnotetext[5]{Also Istituto di Cosmo-Geofisica del C.N.R., Torino,
Italy.}
\footnotetext[6]{Supported by the Commission of the European Communities,
contract ERBCHBICT941234.}
\footnotetext[7]{Supported by CICYT, Spain.}
\footnotetext[8]{Supported by the National Science Foundation of China.}
\footnotetext[9]{Supported by the Danish Natural Science Research Council.}
\footnotetext[10]{Supported by the UK Particle Physics and Astronomy Research
Council.}
\footnotetext[11]{Supported by the US Department of Energy, grant
DE-FG0295-ER40896.}
\footnotetext[12]{Also at CERN, 1211 Geneva 23,Switzerland.}
\footnotetext[13]{Supported by the US Department of Energy, contract
DE-FG05-92ER40742.}
\footnotetext[14]{Supported by the US Department of Energy, contract
DE-FC05-85ER250000.}
\footnotetext[15]{Permanent address: Universitat de Barcelona, 08208 Barcelona,
Spain.}
\footnotetext[16]{Supported by the Bundesministerium f\"ur Bildung,
Wissenschaft, Forschung und Technologie, Fed. Rep. of Germany.}
\footnotetext[17]{Supported by the Direction des Sciences de la
Mati\`ere, C.E.A.}
\footnotetext[18]{Supported by Fonds zur F\"orderung der wissenschaftlichen
Forschung, Austria.}
\footnotetext[19]{Supported by the US Department of Energy,
grant DE-FG03-92ER40689.}
\footnotetext[20]{Now at School of Operations Research and Industrial
Engireering, Cornell University, Ithaca, NY 14853-3801, U.S.A.}
\footnotetext[21]{Now at Schweizerischer Bankverein, Basel, Switzerland.}
%
%
\setlength{\parskip}{\saveparskip}
\setlength{\textheight}{\savetextheight}
\setlength{\topmargin}{\savetopmargin}
\setlength{\textwidth}{\savetextwidth}
\setlength{\oddsidemargin}{\saveoddsidemargin}
\setlength{\topsep}{\savetopsep}
\normalsize
\newpage
\pagestyle{plain}
\setcounter{page}{1}

\section{Introduction}
\label{sec:intro}

 In 1996, a new regime in $\epem$ collisions was entered when LEP 
energies reached and exceeded the $W$ pair production threshold. 
Data were collected with the ALEPH detector at $\sqrt s = 161.3$~GeV 
($10.8~\pbinv$), $\sqrt s = 170.3$~GeV ($1.1~\pbinv$) and  
$\sqrt s = 172.3$~GeV  ($9.6~\pbinv$). 
The increased centre-of-mass energies motivate the direct search for
new physics, in particular for particles predicted by  \susyk\ theories.


Supersymmetry (SUSY)~\cite{SUSYTH} requires the number of degrees 
of freedom associated with the fermionic and bosonic fields 
of the theory to be the same. This is achieved by augmenting 
the ordinary field multiplets with additional fields 
differing by a half unit of spin. The resulting particle 
spectrum contains several new states: gauginos, associated 
with the ordinary gauge bosons; Higgsinos with the Higgs 
bosons; sleptons, sneutrinos and squarks with the ordinary 
matter fermions.
Here, searches for \susyk\ partners of the gauge and Higgs 
bosons are reported,
 while searches at these energies for sleptons~\cite{sleptons}, 
stops~\cite{stops} and Higgs bosons~\cite{Higgs} have been reported previously.
Searches similar to those discussed here have been 
reported by the OPAL collaboration~\cite{opal-172}.

The Minimal Supersymmetric Standard Model (MSSM) is the supersymmetric 
extension of the Standard Model with minimal field content.
Two doublets of complex scalar fields are introduced to give mass to the 
up-like and down-like fermions via the Higgs mechanism. The ratio of the 
two vacuum expectation values is denoted $\tb$ and 
the Higgs mass term is $\mu$.
Soft SUSY 
breaking terms lift the mass degeneracy of ordinary particles and their SUSY 
partners. The scale of these terms should not exceed 
${\sim} 1~\Tcsq$ in  order for \susy\ to remain a solution of the naturalness 
problem. These SUSY breaking terms are: 
gaugino masses $M_1$, $M_2$ and $M_3$, associated to the $U(1)$, $SU(2)$ and 
$SU(3)$ gauge groups, respectively; and mass terms $m_i$ and trilinear couplings  
$A_i$ for the various sfermions. 
The partners of the photon, Z and neutral Higgs bosons 
mix to form 
four mass eigenstates called neutralinos, $\Pchi, \Pchip,
\Pchipp,\Pchippp$, in order of increasing mass. 
Similarly, charged 
gauginos ($\widetilde{\mathrm W}^+$) and Higgsinos ($\widetilde{\mathrm H}^+$)
form charginos, $\chi^{\pm}$ and $\chi^{\pm}_{2}$.
Ordinary particles and \susyk\ particles are distinguished by their 
R-parity, a multiplicative quantum number, which is assumed to be conserved
to ensure lepton and baryon number conservation. As a consequence, 
\susyk\ particles are produced in 
pairs and decay to the Lightest Supersymmetric Particle (LSP), assumed  here 
to be  the lightest neutralino, which is weakly interacting and does not decay,
escaping detection.

The large number of free parameters in the MSSM can be reduced by making 
certain theoretical assumptions. First, the gaugino masses may be assumed to 
unify at the GUT scale, leading, at the electroweak scale, 
to the relation: $M_1=\frac{5}{3}\tan^2\theta_W\ M_2$.
Second, the masses of the sleptons
might also unify at the GUT scale with value $m_0$.  Their masses
at the electroweak scale are derived using the renormalization
group equations~\cite{RGE}, and are an increasing function of $m_0$.  
These assumptions are made for many
of the results presented here.
A special effort is made to interpret the results within a larger 
framework, relaxing the gaugino mass and/or the scalar mass unification 
assumptions.

Given the large value of the top quark mass~\cite{Mtop},  the 
``infrared quasi fixed point scenario''~\cite{fixed_point} favours low 
($\tb {\sim} 1-3$) or high ($\tb {\sim} 30$) values of $\tb$. The various selections 
are optimised  for a value of $\tanb$ equal to $\sqrt 2$, typical of the
low $\tb$ solution. Unless otherwise specified, the results are presented 
for that same value of $\tanb$.
High $\tb$ values tend to give stronger constraints.

With unification of gaugino mass terms, in the region where $\M \gg|\mu|$
the lightest chargino and  neutralino have large Higgsino components; this is 
referred to as the ``Higgsino'' region. Here, the lightest neutralino $\chi$ is 
generally close in mass to the lightest chargino and to the second lightest 
neutralino. Similarly, the region where $|\mu|\gg\M\ $ is referred to as the 
``gaugino'' region; here, $M_\chi \simeq M_{\PCha}/2$. 
In both regions, $M_{\Pchip} > \Mcha$. 
In the region of small negative $\mu$ and low $M_2$, one of the two lightest 
neutralinos has large gaugino components, while the other has large Higgsino 
components. The chargino has sizeable gaugino and Higgsino components. This 
region will be  referred to as the ``mixed'' region.

At LEP, charginos are pair produced by virtual photon or Z exchange in 
the $s$~channel, and sneutrino exchange in the $t$~channel~\cite{chaprod}. 
The $s$  and 
$t$~channels interfere destructively, so that low sneutrino masses lead to 
smaller cross sections. Neutralinos are  produced by $s$-channel Z exchange 
and $t$-channel selectron exchange~\cite{chiprod}. 
Here, the $s$  and $t$~channels interfere 
constructively for most of the parameter space. As a consequence, cross 
sections are usually higher if selectrons are light. 

Charginos decay to a neutralino and a lepton-neutrino or quark-antiquark pair. 
If all sfermions are heavy (large $m_0$), the decay proceeds mainly through 
the exchange of a virtual W. The dominant final state topologies for chargino 
pair production are then hadronic events with missing energy carried away by 
the two neutralino LSP's, called here the four-jet topology (4J), or 
 events with hadrons, an isolated lepton and missing energy (2JL topology). 
Acoplanar lepton pairs (AL topology) are 
also produced, but at a much lower rate. The second lightest neutralino 
\Pchip\ decays to a neutralino and a fermion-antifermion pair. If all 
sfermions are heavy, the decay proceeds mainly through the 
exchange of a virtual Z. The main final state resulting from $\chi\Pchip$ 
production therefore consists of acoplanar jets (AJ), due to the small Z 
leptonic branching ratio. (The $\chi\chi$ final state is invisible.)

Selections for the 4J, 2JL and AJ topologies are designed for chargino 
and neutralino masses close to the kinematic limit for $\chi^+\chi^-$ or 
$\chi\Pchip$ production and are optimised 
for decays dominated by  W* (Z*) exchange and 
for various \dm\ ranges. Here \dm\ is the mass difference between the lightest 
neutralino and the chargino or the second lightest neutralino. 
The signal properties, and hence the background composition and significance,
change dramatically with the mass difference: for low \dm, the phase space for 
decay is small and the signal topology resembles that of
$\epem \rightarrow \epem {\mathrm {f\bar{\mathrm f}}}$ events, while for 
very large \dm, as is the case for massless neutralinos, the signal for 
chargino production is more WW-like. Small mass differences are typical 
of the Higgsino region while large mass differences correspond to 
the gaugino  and mixed regions. 

When sleptons are light, leptonic chargino and neutralino decays are enhanced, 
due to the increased influence of slepton exchange diagrams. 
Since no signal for squarks has been found at the Tevatron~\cite{SquarkSearch},
their masses should be heavy enough to make any influence 
on chargino and neutralino decays negligible. On the 
contrary, sleptons are expected to be lighter 
due to smaller radiative corrections to their 
masses; hence light slepton effects on the leptonic branching 
ratios of charginos and neutralinos can not be neglected.
The dominant topologies are then acoplanar lepton pairs, and the 
selections are based on  those designed for the slepton 
searches~\cite{sleptons}. 
When sleptons are light enough, two-body decays such as 
$\PCha \to l \tilde\nu$ or $\Pchip\to\nu\tilde\nu$ open up, the latter 
leading to an invisible final state. 

In the Higgsino region, $\chi\Pchip$ production is the only relevant 
neutralino process. In the mixed region, the heavier neutralinos 
$\Pchipp$ and $\Pchippp$ can also be produced. Complex topologies arising from 
cascade and radiative decays are covered by a few dedicated searches.

In the various selections, the cuts are adjusted 
according to the ``$\Nbarnf$ prescription''~\cite{nbar}. 
The optimal compromise between signal efficiency and background level
is obtained when the expected $95\%$ C.L. upper limit on the  
cross section is minimised based on Monte Carlo simulations; this optimum 
changes, {\it i.e.},
the cuts become tighter, as the integrated luminosity increases. 
The $\gagahh$ background is handled differently, due to the difficulties to model 
this process  and to simulate the detector response at the level of accuracy 
imposed by its large cross section. In this case, more severe criteria are
applied than would result from the $\Nbarnf$   procedure, as explained in 
Section~\ref{gghh}. For any given chargino and neutralino 
mass combination (or any choice of $M_2$, $\mu$ and $\tanb$) and for any 
given leptonic branching ratio (or any choice of $m_0$), an optimal 
combination of the various selections is chosen, again according to the 
$\Nbarnf$ prescription, counting events that pass any of the combined 
selections to determine efficiencies and contaminations.

In the specific case of large 
slepton masses, the selection efficiency for chargino pair
production depends mostly on the chargino and neutralino masses. Therefore,
results can be presented not only in the MSSM parameter space but also in 
terms of limits on the chargino pair production
cross section as a function of these two masses in a fairly general way. 
A similar statement holds for $\chi\Pchip$ production (neglecting possible 
$\Pchip \to \chi \gamma$ decays at this stage).
To take into account in a consistent way the other neutralino production 
channels, such as $e^+e^- \to \Pchip\Pchipp$, and also to cope with lower 
$m_0$ values, the validity of the MSSM with all the unification conditions 
mentioned earlier is assumed. 
Because of possible large mixing in the stau sector, three-body decays 
involving taus can dominate over those involving other lepton flavours, and 
the two-body decay $\PCha \to {\tilde\tau} \nu$ may open up before 
$\PCha \to l \tilde\nu$. Attention is given to those possibilities.
The dependence of the results on the assumption of universality of scalar 
masses and gaugino masses is investigated.

The outline of this 
paper is as follows: after a brief description of the ALEPH detector in 
Section~\ref{sec:detector}, the various selections are detailed in 
Section~\ref{sec:chaneusel}, the results and their interpretation are 
presented in Section~\ref{sec:results} and conclusions are drawn in 
Section~\ref{sec:conclusions}.

\section{The ALEPH Detector}
\label{sec:detector}

The ALEPH detector is described in detail in~\cite{AlephDetector} and its
performances in~\cite{AlephPerformances}; 
only the features most relevant for the chargino and neutralino analyses 
are given here.

The detector is required to be fully 
operational. At least one of the  
 major triggers for \susy\ searches (total energy triggers, 
single charged electromagnetic and single muon triggers~\cite{AlephDetector}) 
is required to be 
fired. 

Charged  particle tracks are measured by a silicon vertex detector, a drift 
chamber and a large time projection chamber (TPC), immersed in a 1.5 Tesla 
magnetic field provided by a superconducting magnet. A momentum resolution up 
to  $\mathrm \Delta P_{T}/P_{T} = 6 \times 10^{-4} P_T + 0.005\ 
( P_T$ in $\mathrm{GeV}/c)$  can be achieved. 

 The electromagnetic calorimeter (ECAL), a sandwich of lead sheets and 
proportional tubes, is located inside the coil. Highly granular transverse and 
longitudinal measurements of electromagnetic showers are provided by 
projective towers, which are segmented longitudinally in three storeys. The 
achieved energy resolution is 
$\Delta E/E \simeq 18\%/\sqrt E + 0.009$ ($E$ in GeV). The 
ECAL angular coverage extends down to within $10^{\circ}$ from the beam axis.


  The iron return yoke is instrumented as a hadron calorimeter
(HCAL)  consisting of projective towers giving a measurement
of the shower energy.  The pattern of fired streamer tubes provides a
two-dimensional view of the energy deposit, which is useful for identifying
muons.
The HCAL covers polar angles down to $\theta = 8^\circ$. Streamer 
chambers outside of the HCAL (``muon chambers'') are used to tag penetrating 
charged particles.

The calorimetric coverage is extended down to polar angles of 
24 mrad by the luminosity calorimeters LCAL and SICAL.
The low-angle acceptance of SICAL, below 34 mrad, is shadowed by a shielding
mask installed in 1996 to reduce the potentially higher machine background 
at LEP2.

   As the main signal for the processes searched here is missing energy, a good
hermiticity of the detector is essential. 
The ECAL and HCAL cracks are not
aligned, so there are no acceptance holes in ALEPH at 
large polar angles. 
The HCAL covers the gap between the ECAL and the LCAL 
so particles originating from the interaction point passing through 
this gap are detected. The LCAL consists of two half modules on each side of 
the detector, with a small vertical inactive region. This crack is partially 
covered by HCAL.

  The information from the tracking detectors and the calorimeters is 
combined by an energy flow algorithm described in~\cite{AlephPerformances}. 
For each event, a list of energy flow objects (charged particles, photons, 
neutral hadrons, clusters in the luminosity calorimeters) is provided. The 
analyses presented here are based on these objects such that, for example, the
visible mass is the invariant mass of all objects  and the $P_T$ is the 
component transverse to the beam axis of the momentum sum of all objects.

  Lepton identification is described in~\cite{AlephPerformances}. Electrons 
are identified using their specific ionisation in the TPC ($\dedx$) 
and the transverse 
and longitudinal shower shapes in ECAL. Muons are separated from hadrons
by their characteristic penetrating pattern in HCAL 
and the presence of hits in the muon chambers.

\rm
\section{Searches for Charginos and Neutralinos}
\label{sec:chaneusel}

 The various selections for chargino and neutralino searches
are presented in this section.
The chargino analyses in the 4J and 2JL topologies are designed for 
four \dm\ regions: very low (VL) for $\dm \simeq 5~\Gcsq$, low (L) for 
$\dm \simeq 10~\Gcsq$, high (H) ($\dm \simeq M_{\chi^+}/2$) and very high (VH)
($\dm \simeq 80~\Gcsq$). Two neutralino analyses in the AJ topology 
(AJ-L and AJ-H) are designed for \dm\ smaller than or greater than 
$30~\Gcsq$. Additional selections address the acoplanar lepton pair topology 
and the more complex final states encountered in the neutralino searches.

  The 4J-VH and 2JL-VH chargino selections are optimised separately for 
$\sqrt s = 161$~GeV and $\sqrt s = 172$~GeV, due to a large 
increase in the WW cross section, which is the most difficult background for 
this range of \dm. In the other chargino and neutralino selections, a single 
set 
of cuts is  applied for the two energies; this does not degrade the 
performance of the analyses at either energy. 
Although the selections can be optimised for a given $\dm$, in general
they are not optimal when $\dm$ is changed by a small amount.
In order to maintain a smooth transition between the various \dm\ 
regions in the chargino analysis (VH, H, L, VL), each selection is optimised 
on a signal containing an admixture of the nearby \dm\ configurations, 
typically weighting by $50\%$ a configuration at the midpoint 
of the $\dm$ range, $\overline{\dm}$, 
and $25\%$ each at $\dm = \overline{\dm} \pm 20~\Gcsq$.
The optimisation procedure is performed by means of a program that varies
all cuts simultaneously, in order to take correlations between variables
into account.

In the following, the various Monte Carlo samples used for selection 
optimisation are described first. The specific criteria applied to reject the 
\gagahh\ background are addressed next. The searches for charginos are then 
described, followed by the searches for neutralinos. The combinations 
of selections applied in the various cases are presented, and the 
corresponding search efficiencies are given. A discussion of the systematic 
errors follows.

\subsection{Monte Carlo Sample}
The Monte Carlo 
generator DFGT~\cite{DFGT}
is used to simulate the chargino events. 
This generator takes into account the 
spin correlation 
in the production and decay of charginos.
The final states are interfaced for the hadronisation process to
JETSET~7.4~\cite{JETSET} and initial state radiation is included.

  Another widely used generator for \susy\ is SUSYGEN~\cite{SUSYGEN} which 
simulates the production and decays of charginos, neutralinos, sleptons and 
squarks, including cascade decays. It is used for neutralino production and 
as a cross-check for chargino production. Final state radiation is taken into
account using PHOTOS~\cite{PHOTOS}.

  Dilepton Standard Model processes are simulated with KORALZ~\cite{KORALZ}
for $\mu^+\mu^-$ and $\tau^+\tau^-$ and UNIBAB~\cite{UNIBAB} for Bhabha 
production. KORALW~\cite{KORALW} is used to generate W-pair events. 
Leptonic two-photon events are generated with PHOT02~\cite{PHOT02}. 
PYTHIA~\cite{PYTHIA} is used to generate $\gagahh$ events. This generator
is restricted to ``untagged'' events where the outgoing electrons are nearly
undeflected. The PYTHIA sample is complemented by events generated with PHOT02
where an electron is required to be deflected by at least 5~mrad.
The $\gaga$ events are generated with an invariant mass cut of 
$3.5~\Gcsq$.
All other processes ($\qqg$, $\mathrm{ZZ}$, $\mathrm{We}\nu$ and 
$\mathrm{Zee}$) are generated using PYTHIA, with an invariant mass cut 
for the resonance of 0.2~$\Gcsq$ for $\mathrm{ZZ}$ and $\mathrm{We}\nu$ and 
$2~\Gcsq$ for $\mathrm{Zee}$. Here, ``Z'' also includes $\mathrm{Z}^*$ 
and $\gamma^*$ production.
Samples corresponding to at least 20 times the integrated luminosity 
of the data are
generated, except for \gagahh. This process is simulated 
with three times the integrated luminosity of the data,
only for 
161 GeV because of the slow dependence of the event properties on~$\sqrt s$.

\subsection{Rejection of ${\bf \gamma\gamma\rightarrow hadrons}$ }
{\label{gghh}}

The $\gagahh$ background is particularly important for the VL and L mass 
difference regions but may also contaminate other analyses due to its 
large cross-section ($\sim 13$~nb). First the strategy and the variables
used for the rejection of this background are presented, then the specific 
cuts for each chargino and neutralino selection are discussed.

\subsubsection{Strategy}



Transverse momentum is the most natural quantity to reject 
this background. For an ideal detector covering the solid angle above 34 mrad
 with unlimited precision, requiring the event transverse momentum 
$P_T$ to be in 
excess of $3~\pcecm$ should  reject all 
the $\gagahh$ events. Unfortunately, the measurement of the 
visible system is not ideal and fluctuations can induce 
``fake'' $P_T$; hence 
rejection of events with some energy at low angle is needed. 
However, the energy veto 
may be rendered ineffective by inactive regions, such as 
the vertical LCAL crack; new variables, 
discussed in the next section, are therefore needed to detect
these situations.
Furthermore,
the Monte Carlo prediction for this background suffers 
from detector simulation problems 
and from inaccuracy in the simulation of the underlying physics.

Due to these potential problems, energy-based quantities are 
used in conjunction with quantities based on direction measurements to ensure
a better rejection of \gaga\ events. 
Any event that is rejected by only one cut is still counted as a fraction of 
an event in the background estimation, the exact value depending on how far 
away the cut variable for this event is from the cut value, taking into 
account the distribution of the variable as well as its sensitivity to 
reconstruction problems.

\subsubsection{Variables}

  Due to the simple kinematics of \gaga\ events, it is possible to reconstruct
the four-momentum of the outgoing deflected electron (positron) from the 
properties
of the event, assuming that the outgoing positron (electron) is not deflected
at all ($\theta=0^{\circ}$). The scattering angle \thscat\  is defined to 
be the minimum polar angle of the deflected particle for the two hypotheses. 
The pointing angle \thpoint\   is the minimum angle between the calculated 
deflected particle and any energy flow object. 
In \gaga\ events, for large enough \thscat, the outgoing particle should be 
visible. Even if only a fraction of its energy is detected, {\em e.g.} due to 
cracks, \thpoint\ should be small, as the reconstructed particle will 
``point'' to the energy deposits.


  The acoplanarity ($\acop$) is the 
azimuthal angle between the momentum sum of particles
in each hemisphere of an event, defined by a plane perpendicular to the 
thrust axis.
Due to the large missing momentum along the beam axis in \gaga\ events, the 
transverse acoplanarity is introduced. It is calculated by first projecting 
the event onto the plane perpendicular to the beam axis, 
calculating the two-dimensional thrust axis and dividing the event into two 
hemispheres by a plane perpendicular to this thrust axis. The transverse 
acoplanarity $\acopt$ is the angle between the
two hemisphere 
 momentum sums. 

   Since the visible energy is small in \gaga\ events, the kinematics of the 
event can be distorted by minor reconstruction problems. In particular, fake
neutral hadrons are created when the energy deposit of a charged particle in 
the calorimeters is not associated with its reconstructed track. To cope with 
such events, the fraction of the visible energy  carried by neutral hadrons 
($F_{NH}=E_{NH}/\evis$) as well as the $P_T$ of the event excluding neutral 
hadrons from its calculation (\ptnonh) are used.

\subsubsection{Common cuts}

Monte Carlo \gaga\ events are generated with an invariant mass of the 
hadronic system greater than~$3.5~\Gcsq$; thus, in all analyses, the visible 
mass is required to be greater than~$4~\Gcsq$.

With the required set of triggers, trigger inefficiency is not negligible for 
\gaga\ events nor for signal events in the very low \dm\ region, but 
is small for events that would have been selected by any of the selections 
described hereafter. The most important triggers for \gaga\ events rely on the 
energy measurement in ECAL. It has been checked that the Monte Carlo 
reproduces well the energy distribution except for the threshold behaviour; in 
order to avoid that potentially dangerous region, higher trigger thresholds 
are applied offline on data and Monte Carlo events.

 The cuts designed to eliminate the \gagahh\ background for the very high, 
high and low mass difference chargino analyses and for neutralino analyses are 
detailed here and summarised in Table~\ref{ggcuts}. 
Specific analyses, described in 
Section~\ref{chasel}, are designed in the very low $\dm$ case, where 
the main background is \gagahh. 

\subsubsection{Rejection of ${\bf \gamma\gamma\rightarrow hadrons}$ 
in the chargino selections}

\paragraph{Four Jet Topology (4J)}
   For the 4J-VH and 4J-H selections, the number of good charged tracks 
is required to 
exceed six. The event $thrust$ must not exceed 0.9. The 
$P_T$ is required  to be larger than~$5\%\sqrt s$ or~$7.5\pcecm$ if the 
azimuthal angle of the missing momentum \phmiss\  is within~$15^{\circ}$ of 
the vertical plane. Although the actual size of the LCAL vertical crack is 
much smaller than~$15^{\circ}$, the angular region that defines it must take 
into account the resolution on the missing momentum direction. 
The transverse acoplanarity must be less than~$175^{\circ}$. The energy 
detected within~$12^{\circ}$ of the beam axis, $E_{12}$, must be lower 
than~$5\pcecm$ and \thscat\ must be greater than~$15^{\circ}$ or \thpoint\ 
must be greater than~$5^{\circ}$. The missing 
momentum is required to point at least~$18.2^{\circ}$ away from the beam axis 
($\cmiss<0.95$). To reject events with fake neutral 
hadrons, $F_{NH}$ must be lower than~$30\%$. 
This cut is relaxed to~$45\%$ if \ptnonh\ is greater than~$3\pcecm$.
Finally, to avoid large angle tagged \gaga\ events, the energy 
of the most energetic lepton of the event must not exceed~$20\pcecm$. 

  For the 4J-L chargino analysis, at least four good tracks must be 
reconstructed. All cuts defined above are applied, with the exception of the 
$thrust$ cut. In addition, $E_{12}$ must be zero and the \thpoint\ cut is 
tightened to~$10^{\circ}$. Finally, the energy in a~$30^{\circ}$ azimuthal 
wedge around the direction of the missing momentum (\ewedge) must not 
exceed~$1.5\pcecm$.

\paragraph{Two Jets and Lepton Topology (2JL)}

  The rejection of the \gaga\ background in the 2JL selections is easier due
to the presence of an identified lepton. An electron (muon) of at least~2~GeV 
(2.5~GeV) must be  identified. The number of charged tracks (including 
the lepton) must be at least three and $\acopt$ lower than~$175^{\circ}$. 
\thscat\ must be greater than~$15^{\circ}$ or \thpoint\ greater 
than~$5^{\circ}$. The neutral hadron energy fraction is required to be less 
than~$45\%$. 
In the 2JL-VH and 2JL-H selections, the $P_T$ and $E_{12}$ cuts are the same 
as for the 4J-H channel. In the 2JL-H selection, the energy of the most 
energetic lepton must be less than~$20\pcecm$. This cut would greatly  reduce 
the efficiency for the 2JL-VH selection, and is relaxed to~$30\pcecm$; in 
addition, the missing mass $\wmiss$ is required to be greater than~$25\pcecm$. 
In the 2JL-L analysis, the $P_T$ must exceed~$2.5\pcecm$, $E_{12}$ must be 
zero and $\cmiss<0.95$. The energy in a cone of~$30^{\circ}$ around the most 
energetic lepton in the event (\elcone) is calculated, and  \elcone\ or 
\ewedge\ must be lower than~$1\pcecm$.


  No events from the  various \gagahh\ Monte Carlo samples survive the 4J
and 2JL selection cuts 
and none are rejected by only one cut. For the double rejection requirement, 
the $P_T$ cut is reinforced by the $\acopt$ cut and by the $F_{NH}$ and 
\ptnonh\ cuts in the case of fake neutral hadrons. The $E_{12}$ and pointing 
(\thscat\ and \thpoint) cuts reinforce each other (see Table~\ref{ggcuts}).

\subsubsection{Rejection of ${\bf \gamma\gamma\rightarrow hadrons}$ in 
the neutralino AJ selections}

     
In the neutralino AJ selections, the cuts on $\cmiss$, $E_{12}$, \elept, 
\thpoint\ and \thscat\ are the same as in the 4J-H analysis. The neutral 
hadron energy is required to be less than 45$\%$ of the visible energy.
In the AJ-L analysis,  at least four good tracks must be reconstructed in the 
event and the $P_{T}$ is required to exceed~$3\pcecm$ ($4.5\pcecm\ $ if 
\phmiss\  is within 15$^{\circ}$ of the vertical plane).
The transverse acoplanarity is required to 
be smaller than~120$^{\circ}$. $E_{12}$ and \ewedge\ must be equal to zero.
The $P_T$ is required to be larger than~40$\%$ of the visible energy, and 
$F_{NH}$ less than~30$\%$ unless \ptnonh\  is greater than~$1.8\pcecm$. 
In the AJ-H analysis, the number of good tracks must be larger than six and 
the acoplanarity and transverse acoplanarity must be smaller 
than~175$^{\circ}$. The $P_T$ is required to exceed $5\pcecm$ 
($7.5\pcecm$ if \phmiss\ is within~$15^{\circ}$ of the vertical plane) and 
also to be larger than~20$\%$ of the visible energy. The cuts on  $F_{NH}$ and 
$E_{12}$ are the same as in the 4J-H analysis. Finally the fraction of visible 
energy within~$30^{\circ}$ of the beam axis $\efwd$ is required to be less 
than~70$\%$, and \ewedge\ less than~$7.5\pcecm$.

  All Monte Carlo $\gagahh$ events are eliminated by at least two cuts 
in the AJ-H analysis. In the AJ-L analysis,
the \gagahh\ background is estimated to be ${\simeq} 20$~fb 
from the distance to the cut values of a few singly cut events.

\begin{table}
\begin{center}
{\small 
\begin{tabular}{|c||c|c|c|}
\hline
 \multicolumn{4}{|c|}{Chargino - 4J} \\ 
\hline
$\dm$ range & \multicolumn{2}{|c|}{VH,H} & L \\
\hline
\hline
$\mvis$ & \multicolumn{3}{|c|}{$>4$ GeV/c$^2$ and trigger conditions} \\ 
\hline
$\nch$ & \multicolumn{2}{|c|}{$\geq 7$}  & $\geq 4$ \\ 
\hline
$\pt \geq \fu\rts$ / $\fd\rts^\dagger$ & 
 \multicolumn{3}{|c|}{$\fu =  5\%$, $\fd = 7.5\%$} \\ 
\hline
$\act$ & \multicolumn{3}{|c|}{$<175^\circ$} \\ 
\hline
$\elow$ & \multicolumn{2}{|c|}{$<5\%\sqrt{s}$} & $=0$ \\
\hline
$\tsc > \thu$ or $\tpt > \thd$ 
        & \multicolumn{2}{|c|}{$\thu=15^\circ,\thd=5^\circ$} & 
       $\thu=15^\circ,\thd=10^\circ$ \\
\hline
$| \ctmis |$ & \multicolumn{3}{|c|}{$<0.95$} \\
\hline
$\fnh$ & 
 \multicolumn{3}{|c|}{$< 45\%$} \\  
\hline
$\fnh < \fu$ or $\ptnh > \fd $ & 
 \multicolumn{3}{|c|}{$\fu = 30\%, \fd = 3\pcecm$}  \\ 
\hline
$\elep$ & \multicolumn{3}{|c|}{$<20\%\sqrt{s}$} \\
\hline
Thrust & \multicolumn{2}{|c|}{$<0.9$} &  \\
\hline
$\ewed$ & \multicolumn{2}{|c|}{} & $<1.5\%\sqrt{s}$  \\
\hline
\multicolumn{4}{c}{} \\ 
\hline
 \multicolumn{4}{|c|}{Chargino - 2JL} \\ 
\hline
$\dm$ range & VH & H & L \\
\hline
\hline
$\mvis$ & \multicolumn{3}{|c|}{$>4$ GeV/c$^2$ and trigger conditions} \\ 
\hline
$\nch$ & \multicolumn{3}{|c|}{$\geq 3$} \\ 
\hline
 identified $e/\mu$ & \multicolumn{3}{|c|}{$\geq 1$} \\ 
\hline
$\pt \geq \fu\rts$ / $\fd\rts^\dagger$ & 
      \multicolumn{2}{|c|}{$\fu=5\%,\ \fd=7.5\%$} & $\fu=2.5\%,\ \fd=2.5\%$\\
\hline
$\act$ & \multicolumn{3}{|c|}{$<175^\circ$} \\ 
\hline
$\elow$ & \multicolumn{2}{|c|}{$<5\%\sqrt{s}$} & $=0$ \\
\hline
$\tsc > \thu$ or $\tpt > \thd$ 
        & \multicolumn{3}{|c|}{$\thu=15^\circ,\thd=5^\circ$} \\ 
\hline
$| \ctmis |$ & & & $<0.95$ \\
\hline
$\fnh$ & 
 \multicolumn{3}{|c|}{$< 45\%$} \\  
\hline
$\ensuremath{min}(\ewed,\ewedl)$ & & & $<1\%\sqrt{s}$ \\ 
\hline
$\elep$ & $<30\%\sqrt{s}$ & \multicolumn{2}{|c|}{$<20\%\sqrt{s}$} \\
\hline
$\mmiss$ & $>25\%\sqrt{s}$ & & \\
\hline
\multicolumn{4}{c}{} \\ 
\hline
\multicolumn{4}{|c|}{Neutralino - AJ} \\ 
\hline
  $\dm$ range & \multicolumn{2}{|c|}{H} & L \\
\hline
\hline
$\mvis$ & \multicolumn{3}{|c|}{$>4$ GeV/c$^2$ and trigger conditions} \\ 
\hline
$\nch$ & \multicolumn{2}{|c|}{$\geq 7$}  & $\geq 4$ \\ 
\hline
$\pt \geq \fu\rts$ / $\fd\rts^\dagger$ & 
      \multicolumn{2}{|c|}{$\fu=5\%,\ \fd=7.5\%$} & $\fu=3\%,\ \fd=4.5\%$\\
\hline
$\pt/\evis$ & \multicolumn{2}{|c|}{$>20\%$} & $>40\%$ \\
\hline
$\act$ & \multicolumn{2}{|c|}{$<170^\circ$} & $<120^\circ$ \\ 
\hline
$\aco$ & \multicolumn{2}{|c|}{$<170^\circ$} &  \\ 
\hline
$\elow$ & \multicolumn{2}{|c|}{$<5\%\sqrt{s}$} & $=0$ \\
\hline
$\tsc > \thu$ or $\tpt > \thd$ 
        & \multicolumn{3}{|c|}{$\thu=15^\circ,\thd=5^\circ$} \\
\hline
$| \ctmis |$ & \multicolumn{3}{|c|}{$<0.95$} \\
\hline
$\fnh$ & 
 \multicolumn{3}{|c|}{$< 45\%$} \\  
\hline
$\fnh < \fu$ or $\ptnh > \fd $ & 
 \multicolumn{3}{|c|}{$\fu = 30\%$} \\ 
\cline{2-4}
    &  \multicolumn{2}{|c|}{$\fd=3\pcecm$} & $\fd=1.8\pcecm$ \\
\hline
$\elep$ & \multicolumn{3}{|c|}{$<20\%\sqrt{s}$} \\
\hline
$\ewed$ & \multicolumn{2}{|c|}{$<7.5\%\sqrt{s}$} & $=0$  \\
\hline
$\efwd/\evis$ & \multicolumn{2}{|c|}{$<70\%$} &  \\
\hline
\end{tabular}
}
\end{center}
\caption{\em Selection cuts against $\gamma\gamma\to hadrons$.  
 The $^\dagger$ indicates that the cut is applied 
when the azimuthal angle of the missing momentum is 
within $15^{\circ}$ of the vertical plane.
}
\label{ggcuts}
\end{table}

\subsection{Chargino selections}
\label{chasel}

  There are three topologies for chargino searches: a totally  hadronic 
topology (4J), a topology with a lepton and jets (2JL) and a topology of 
two acoplanar leptons (AL), not necessarily of the same flavour. Several 
analyses are employed in each topology 
to provide sensitivity over a 
large range of mass differences (very high, high, low and very low \dm).
The criteria for the rejection of the \gagahh\ background have been described
in the previous Section.

\paragraph{Four-Jet Topology (4J)}

\begin{figure}
\begin{center}
\begin{tabular}{cc}
\mbox{\epsfig{file=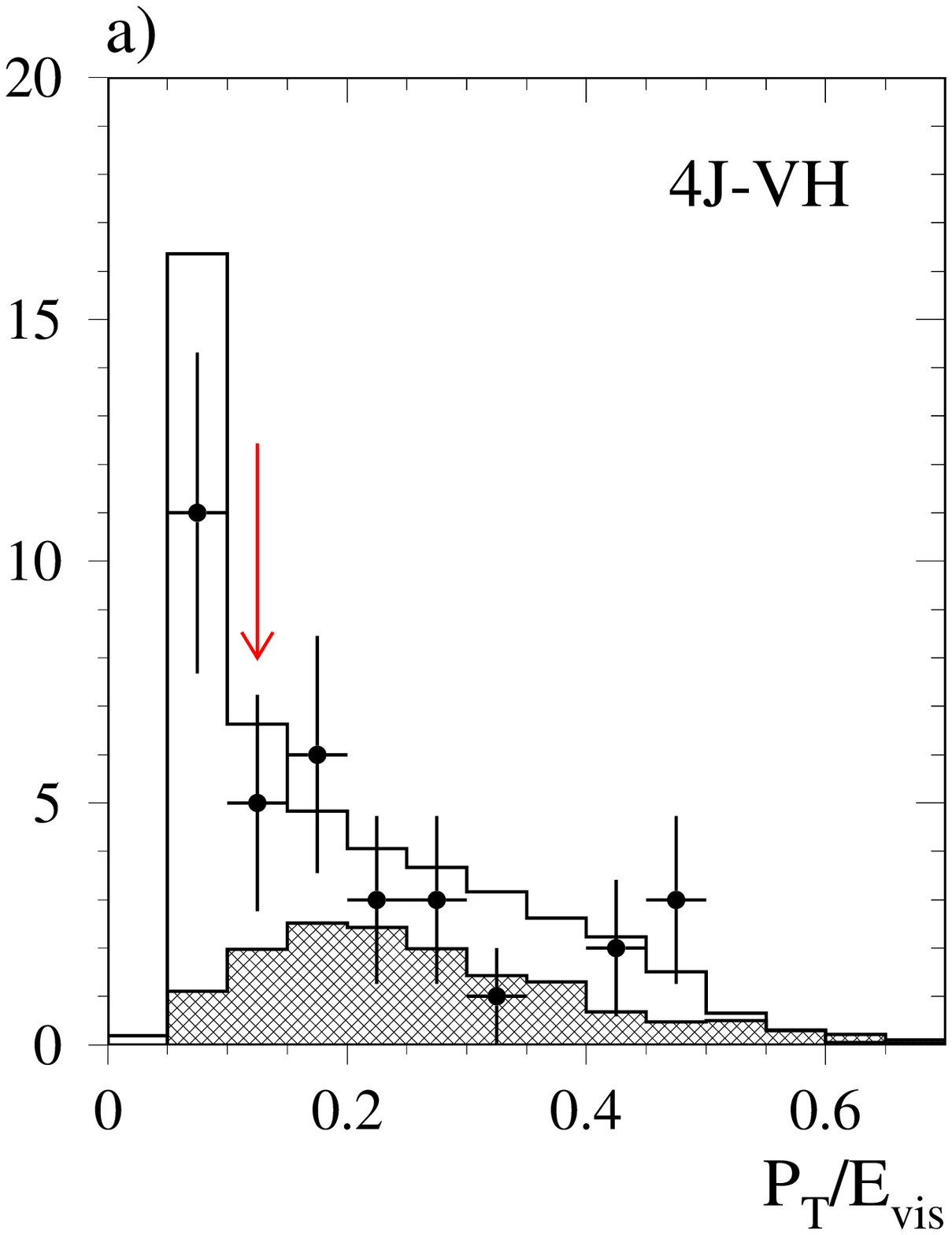,height=8cm}} &
\mbox{\epsfig{file=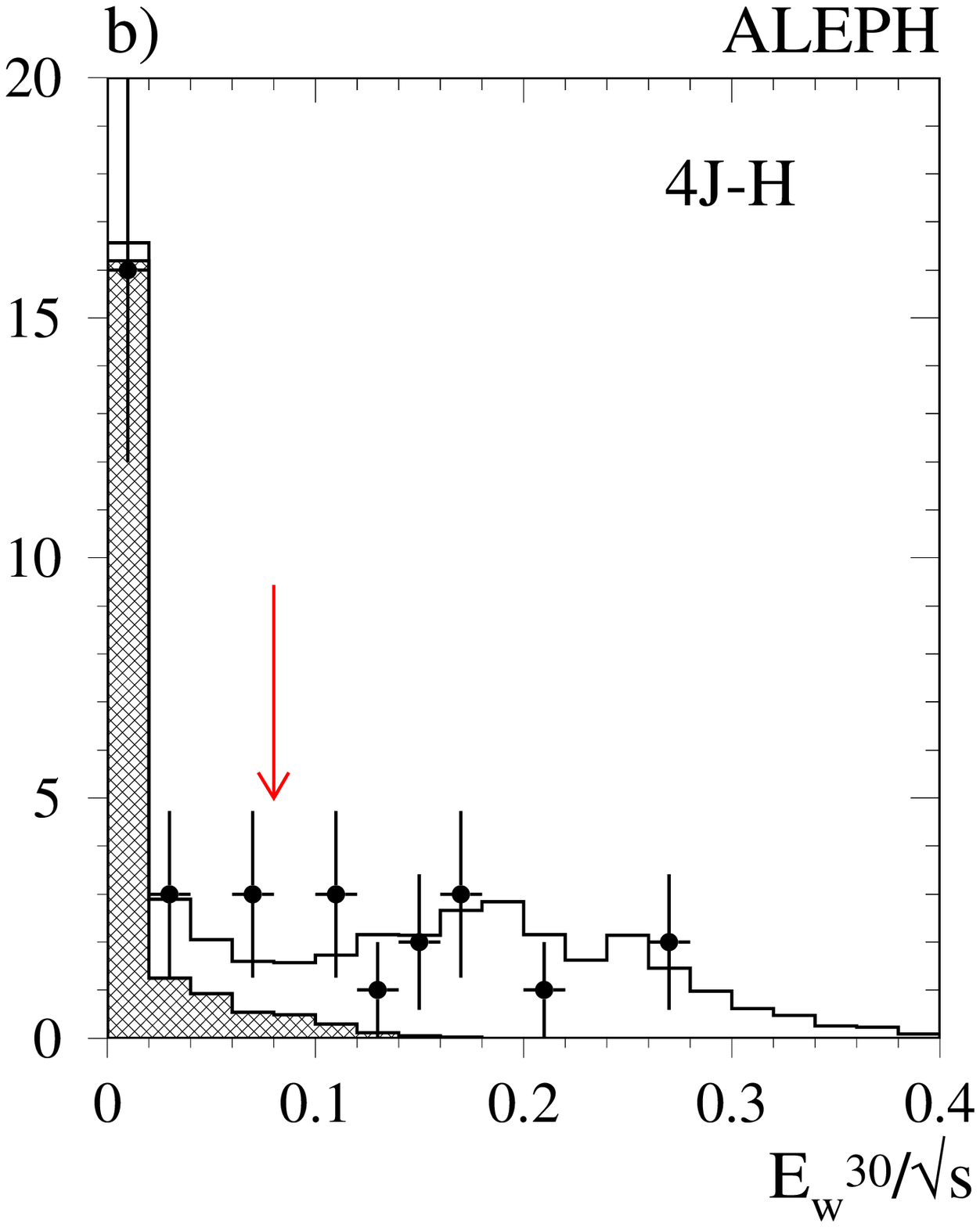,height=8cm}} \\
\mbox{\epsfig{file=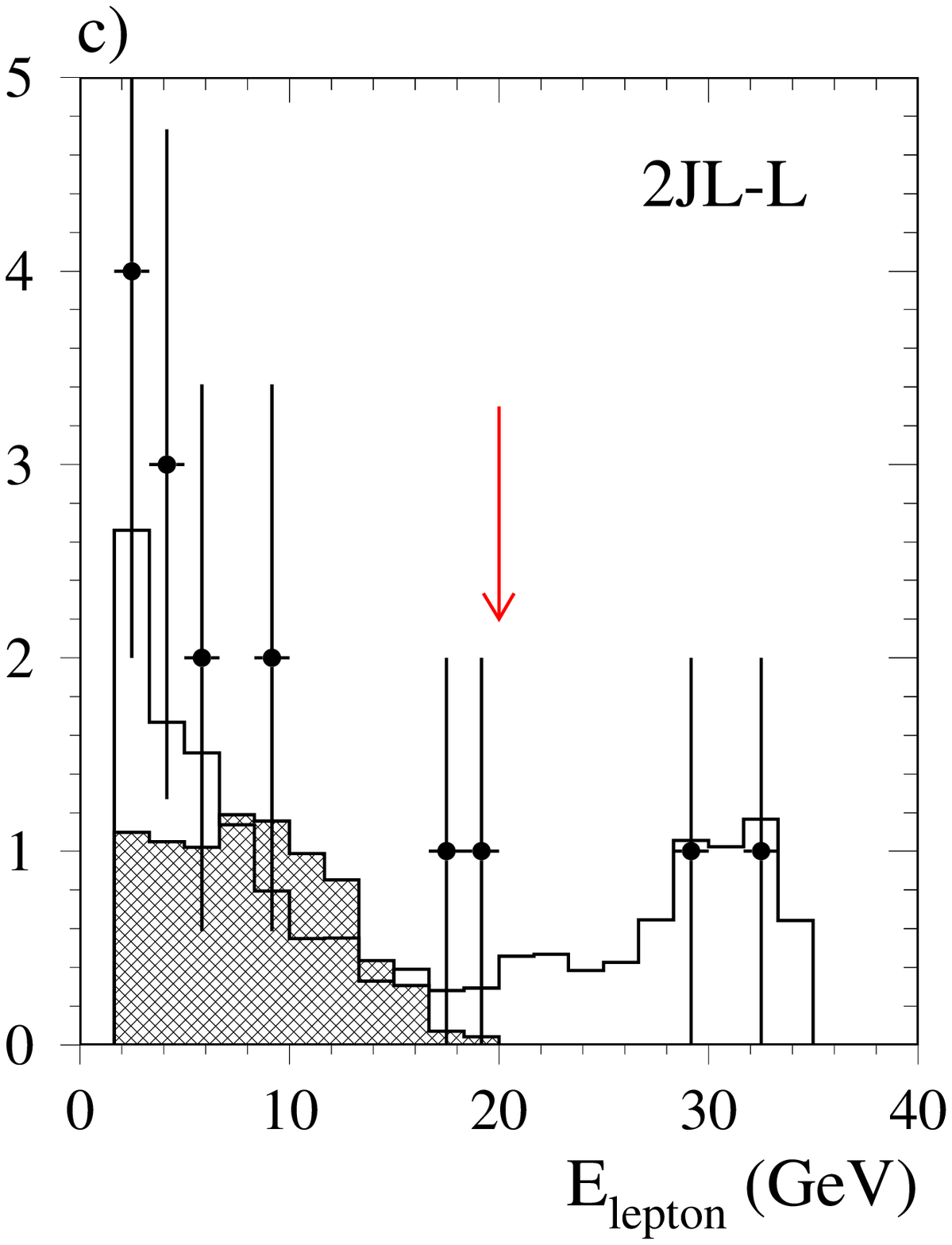,height=8cm}} &
\mbox{\epsfig{file=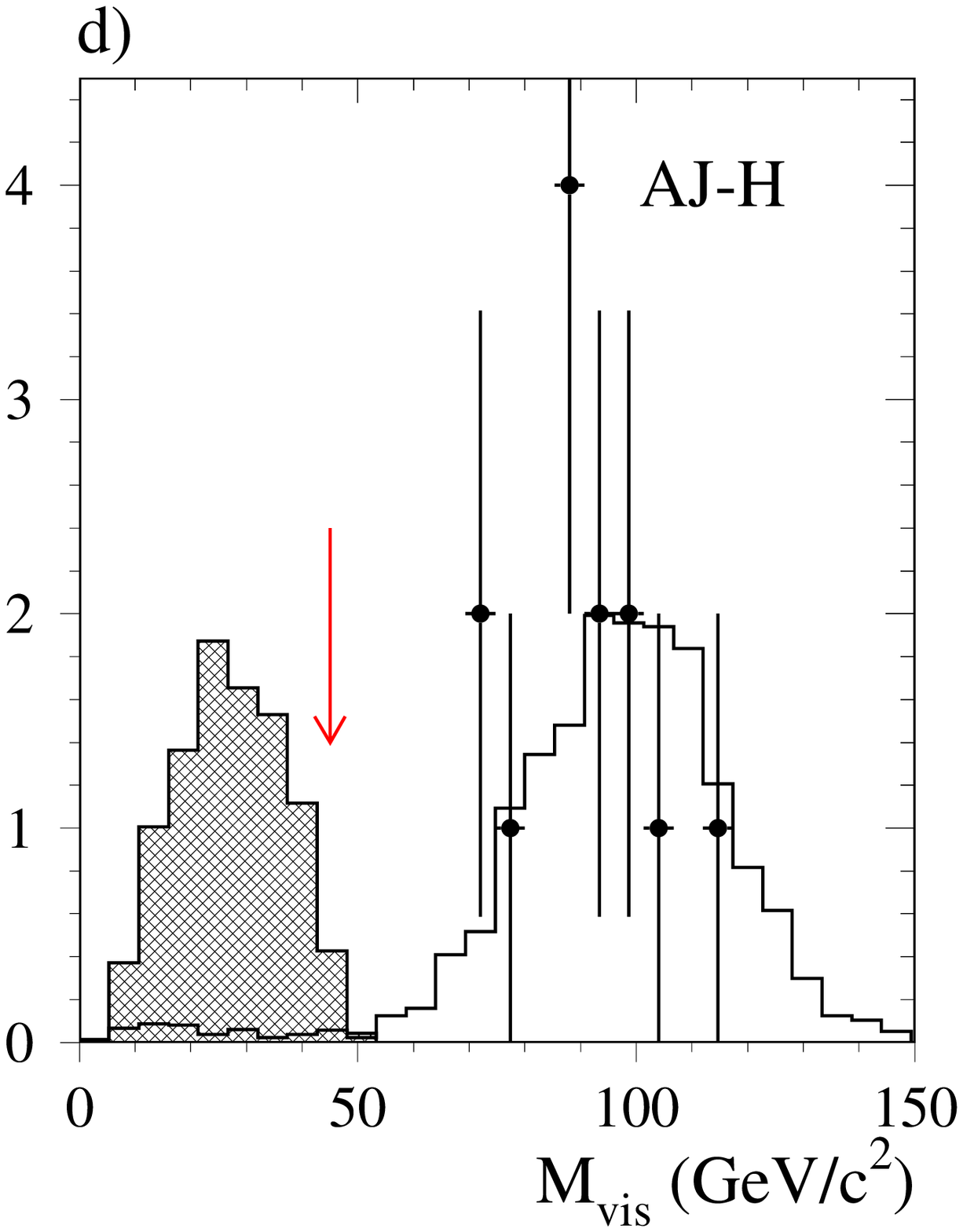,height=8cm}}
\end{tabular}
\caption{\em  
Distributions after $\gagahh$ rejection cuts, for $\rts = 172$~GeV, of
a) $P_T/\evis$ for the 4J-VH selection,
b) $\ewedge$ for the 4J-H selection, c) $\elept$
for the 2JL-L selection and d) $\mvis$ for the AJ-H selection. Points with
error bars represent the data; the solid histogram is the background 
Monte Carlo;
the shaded histogram represents a signal configuration typical of each analysis
($\Mcha-\Mchi (\Gcsq)) = $ a) 85-05 b) 85-45 c) 85-65 and d) ($M_{\Pchip} - \Mchi$) = 
95-60. 
Background MC and signal are normalised to the
integrated luminosity of the data. Signal cross-sections were determined
for $\tb=\sqrt 2$ and for large slepton and sneutrino masses.  The location
of the cut is indicated with an arrow.
\label{fig:distrib} }
\end{center}
\end{figure}

  The most important backgrounds for 4J-VH and 4J-H are $\qqg$ and WW 
production. The missing transverse momentum $P_T$, the transverse 
imbalance $P_T/\evis$ (see Fig.~\ref{fig:distrib}a), 
the visible mass $\mvis$ 
 and the acoplanarity are
used to reduce these backgrounds. No explicit  reconstruction of four jets 
is made
but the event should be spherical (using the $thrust$ or the inverse-boost: 
$\invb = \sqrt{\frac{1}{2}(1/\gamma_1^2+1/\gamma_2^2)}$, where 
$\gamma_i=E_i/m_i$ for each hemisphere of the event). The missing momentum 
should be isolated, as quantified by $\ewedge$ (see Fig.~\ref{fig:distrib}b). 
A veto on a high energy lepton reduces the WW 
background when one of the W's decays to $\mathrm{e} (\mu) \nu$ and an upper cut 
on the  missing mass reduces the WW background when one of the W's decays to 
$\tau\nu$. To further reduce this background, a tau jet is searched for using 
the JADE algorithm with a $y_{cut}$ of $0.001$. The W 
mass ($M_W$) is computed as the mass of the hadronic system excluding the tau 
jet and $\alpha_{23}$ is  defined as the smallest angle between the tau jet 
and the other jets. In the 4J-H and 4J-L analyses, the remaining $q\bar q$ 
radiative return background is reduced by vetoing events with an 
 isolated photon with energy greater than 10~GeV. A photon is 
isolated if no particle is detected 
in a cone of $30^\circ$ half angle around its direction, excluding an inner 
cone of $5^\circ$ half angle. The cuts for the 4J-VH, 4J-H and 4J-L selections 
are listed in Table~\ref{FJcuts}.

\paragraph{Two Jets and Lepton Topology (2JL)}

   The characteristic signature of the 2JL channel is the presence of an 
energetic isolated lepton (see Fig.~\ref{fig:distrib}c).
Here, lepton refers to $\mathrm{e}$ or $\mu$, where electrons are 
identified using 
ECAL information. Cuts on the missing mass and the hadronic mass, $\whad$, 
(the mass of the event, excluding the most energetic identified lepton) 
and $\ewedge$ 
reduce the $q\bar q$ and WW backgrounds. The 2JL-L selection is 
sensitive to the  $\tau^+\tau^-$ and $\gaga\to\tau^+\tau^-$ backgrounds, which
are reduced by means of the hadronic mass and the acoplanarity. The cuts are listed 
in  Table~\ref{FJcuts}. 

\paragraph{Very Low $\dm$ Selections}

   The selections for the very low mass difference chargino signal are 
designed to reject the $\gaga$ background, mostly $\gagahh$ and 
$\gaga\to\tau^+\tau^-$. 
In both 4J and 2JL selections, variables similar to those used for the 
other chargino analyses are employed, such as the  missing transverse 
momentum, the transverse imbalance and the 
transverse acoplanarity.
No energy should be detected at low angle. 
In the spirit of Section~\ref{gghh}, 
energy-based variables are complemented by direction-based variables such as 
$\cmiss$, $\thscat$ and $\thpoint$. Events with fake neutral hadrons are 
removed with $F_{NH}$ and \ptnonh, as can be seen in Table~\ref{FJcuts}. 
As the main potential background in the 2JL-VL 
selection comes from events with a misidentified hadron, cuts to identify 
leptons are tighter than in the other  2JL analyses and also 
use the $\dedx$ 
information from the TPC to identify
signal electrons which typically have lower momentum than those targeted by
the L, H, and VH selections. 
The momentum is required to be in excess of $1~\gevc$ for electrons and
$2.5~\gevc$ for muons.
The estimated $\gagahh$ contamination is $\simeq 60$~fb in 4J-VL and 
negligible in 2JL-VL.

\paragraph{Acoplanar Leptons Topology (AL)}

   The acoplanar lepton selection is similar to the ALEPH selectron and smuon
selections~\cite{sleptons}, with the exception that no lepton
identification is required and events with four tracks are accepted, where 
three tracks are hypothesized as arising from a tau decay if their invariant
mass is lower than $1.5\ \mathrm{GeV}/c^2$.
The cuts against WW background, specifically the requirement
 on the energy of the tracks ($\elepu$ and $\elepd$, 
where $\elepd < \elepu$), are optimised for 
the two chargino decay scenarios leading to an acoplanar lepton signature: 
$\PCha \rightarrow \Pchi \ell \nu$ (three-body decay: AL-3) and 
$\PCha \rightarrow \PSnu \ell $ (two-body decay: AL-2). 
Typical cut values  for $\rts = 172$~GeV are $\elepu < 30$~GeV  
and $\elepd < 26$~GeV for the AL-3 selection. 
Similar values of these cuts are applied in the AL-2 selection when
$\mcha - \msnu < 20~\Gcsq$.

For two body decays, the irreducible background from W pair
production is subtracted.
The optimisation of the upper cut on the energy of the leptons in the
two-body decay selection takes this
 into account, and the cut applied depends on the mass
difference between the chargino and sneutrino.
The 
selection for small mass differences described in Ref.~\cite{sleptons} is also
applied in the analysis of the $\PCha \rightarrow \PSnu \ell $ topology 
(AL-VL), requiring a pair of identified leptons,
not necessarily of the same flavor.
This analysis is also applied to the data taken at $\rts = 130$ and $136$~GeV, 
giving similar efficiencies and background as at $\rts = 161$ and $172$~GeV.

\vspace{1cm}

   The background level and typical efficiency for each of the selections 
described above are listed in  Table~\ref{EffBkgChar}. 
To optimally search for a given  chargino signal, some of the selections 
are combined as described in Section~\ref{sec:combination}.

\begin{table}
{\small
\begin{center}
\begin{tabular}{|c||c|c|c|c|}
\hline
\multicolumn{5}{|c|}{Chargino - 4J} \\ 
\hline
  $\dm$ range & VH - 161 GeV & VH - 172 GeV  & H & L \\
\hline
\hline
 anti-$\gamma\gamma$ cuts & \multicolumn{4}{|c|}{Yes} \\
\hline
$\nch$ & $\geq 24$  & $\geq 26$ & \multicolumn{2}{|c|}{} \\ 
\hline
$\pt$ & $>15$ GeV/c  & $>10$ GeV/c & \multicolumn{2}{|c|}{} \\ 
\hline
$\pt/\evis$ & $>12.5\%$ & $>10\%$ & \multicolumn{2}{|c|}{} \\
\hline
$\mvis$ & \multicolumn{2}{|c|}{$<160~\Gcsq$} &$<70~\Gcsq$&$<60~\Gcsq$ \\ 
\hline
$\aco$ & $<165^\circ$ & $<175^\circ$ & \multicolumn{2}{|c|}{}  \\ 
\hline
$\invb$ & \multicolumn{2}{|c|}{$>0.4$} & $>0.3$ & $>0.25$ \\ 
\hline
$\ewed$ & \multicolumn{2}{|c|}{$<10\%\sqrt{s}$} & $<8\%\sqrt{s}$ &   \\
\hline
$\elep$ & $<10$ GeV & $<15$ GeV & & $<20$ GeV \\
\hline
$\mmiss$ & $<60~\Gcsq$ & $<70~\Gcsq$ & & $>100~\Gcsq$ \\
\hline
$\mmw > M_1$ or $\adt > \aref$ & \multicolumn{2}{|c|}{$M_1 = 90~\Gcsq$} 
& \multicolumn{2}{|c|}{}  \\ 
\cline{2-3}
        & $\aref=140^\circ$ & $\aref=80^\circ$ & \multicolumn{2}{|c|}{} \\
\hline 
 Thrust & & & $<0.85$ & $<0.925$ \\
\hline
Isolated $\gamma$ & \multicolumn{2}{|c|}{} & \multicolumn{2}{|c|}{=0}  \\
\hline
\multicolumn{5}{c}{} \\ 
\hline
 \multicolumn{5}{|c|}{Chargino - 2JL} \\ 
\hline
  $\dm$ range & VH - 161 GeV & VH - 172 GeV  & H & L \\
\hline
\hline
 anti-$\gamma\gamma$ cuts & \multicolumn{4}{|c|}{Yes} \\
\hline
$\nch$ & \multicolumn{3}{|c|}{$\geq 7$} & \\ 
\hline
$\elep$ & $>12.5$ GeV & $\in[10,40]$ GeV & $>2.5\%\sqrt{s}$ & $<20$ GeV \\
\hline
$\ewedl < \elref$ and $\ewed < \ewref$ 
        & Yes & Yes & if $\ewedl>0$ & No \\
$(\elref,\ewref)\Rightarrow$ & $(5\%,20\%)\sqrt{s}$ & 
          $(2.5\%,20\%)\sqrt{s}$  & 
          $(20\%,4\%)\sqrt{s}$  &  \\
\hline
$\mmiss$ & $>50~\Gcsq$ & $>55~\Gcsq$ & $>50~\Gcsq$ & $>120~\Gcsq$ \\
\hline
$\whad$ & $<70~\Gcsq$  & $<65~\Gcsq$  & $\in[5,45]~\Gcsq$ &$>1.5~\Gcsq$\\
\hline 
 Thrust & & &  & $<0.95$ \\
\hline
$\act$ & & & & $<170^\circ$  \\ 
       & & & & $(<150^\circ$ if $\nch\leq 4)$  \\ 
\hline
Isolated $\gamma$ & & & $=0$ & \\
\hline
\multicolumn{5}{c}{} \\ 
\hline
 \multicolumn{5}{|c|}{Chargino - VL} \\ 
\hline
  Topology & \multicolumn{2}{|c|}{4J} & \multicolumn{2}{|c|}{2JL} \\
\hline
\hline
 $\mvis$ & \multicolumn{4}{|c|}{$>4~\Gcsq$ and trigger conditions} \\
\hline
 $\mmiss$ & \multicolumn{4}{|c|}{$>140~\Gcsq$} \\
\hline
 $\ewed,\elow$ & \multicolumn{4}{|c|}{$=0$ GeV} \\
\hline
 $\tpt$ & \multicolumn{4}{|c|}{$>10^\circ$} \\
\hline
 $| \ctmis |$ & \multicolumn{4}{|c|}{$<0.8$} \\
\hline
 $\nch$ & \multicolumn{2}{|c|}{$\geq 4$} & \multicolumn{2}{|c|}{$\geq 3$} \\
\hline
 $\pt$ & \multicolumn{2}{|c|}{$>\ensuremath{max}(2.5\%\sqrt{s},40\%\evis)$} & 
        \multicolumn{2}{|c|}{$>25\%\evis$} \\
\hline
 $\act$ & \multicolumn{2}{|c|}{$<125^\circ$} 
        & \multicolumn{2}{|c|}{$<160^\circ$,$<150^\circ$ if $\nch=4$} \\
        & \multicolumn{2}{|c|}{} 
        & \multicolumn{2}{|c|}{$<110^\circ$ if $\nch=3$ and $| \ctmis |>0.7$} \\
\hline
 Thrust & \multicolumn{2}{|c|}{$<0.95$} & \multicolumn{2}{|c|}{$0.9$} \\
\hline
 $\fnh$ & \multicolumn{2}{|c|}{$=0$ or $\ptnh>2\%\sqrt{s}$} & 
          \multicolumn{2}{|c|}{$<0.4$} \\
\hline
 identified $e/\mu$ & \multicolumn{2}{|c|}{} & \multicolumn{2}{|c|}{$\geq 1$} \\
\hline
 $\ewedl$ & \multicolumn{2}{|c|}{} & \multicolumn{2}{|c|}{$< 5$ GeV} \\
\hline
 $\tsc$ & \multicolumn{2}{|c|}{} & 
          \multicolumn{2}{|c|}{$>2^\circ$ if $\ptlep<2~\gevc$} \\
\hline
 $\whad$ & \multicolumn{2}{|c|}{} & 
          \multicolumn{2}{|c|}{$<10~\Gcsq$} \\
\hline
\end{tabular}
\end{center}
\caption{\em Cut values for the chargino analyses.
The VH selections are divided into cuts used at 
$\rts$= 161~GeV and 172~GeV. Cuts against $\gagahh$ are listed in 
Table~\ref{ggcuts} for the L, H, and VH selections; all cuts for the VL 
selections are listed here. 
\label{FJcuts} }
}
\end{table}

\normalsize

\subsection{Neutralino selections}

\label{neusel}

The neutralino analysis is optimised according to the various production 
and decay topologies in the  different regions of the MSSM parameter space. 
The neutralino topologies can be summarised as follows. In the Higgsino region 
the only accessible channel is  $\Pchi\Pchip$ 
production and the signature 
consists of acoplanar jets. In the mixed region, heavier neutralinos ($\Pchipp,
\Pchippp$) are also produced and give rise to cascade decays. For large 
slepton masses (large $m_0$) the final states are mainly multi-hadronic. 
For light slepton masses (small $m_0$) the leptonic branching ratios are 
enhanced, which gives rise to events containing several leptons in the final 
state. Furthermore, there are parameter configurations (in the mixed region 
for low \tanb, when the $\chi\chi'$ mass difference is small) for which  the 
branching ratio for $\chi{'}\rightarrow\chi\gamma$ is large;  therefore
the events often contain isolated photons. 

In other regions, neutralino decays to a neutral Higgs 
boson ($h$) may play a role, 
depending on the assumptions made for the Higgs sector. The efficiency for 
$\chi^j\to\chi h$ is greater than the efficiency for 
$\chi^j\to\chi \mathrm Z^{*}$, in part due to the $\Pnu \PaSnu$ decays of the 
$\mathrm Z^{*}$.  
Conservatively, the branching ratios of decays to 
Higgs bosons are set to zero.
         
     Four different analyses are used
     to cope with the various decay modes throughout the
     parameter space.  The efficiencies and expected background levels for these
analyses are summarised in Table~\ref{EffBkgChar}.

\paragraph{Acoplanar Jets Topology (AJ)}

Two analyses are employed for this topology: the AJ-H analysis, 
optimised for large mass differences between \Pchi\ and $\Pchip$,
and the AJ-L 
analysis, optimised to complement the AJ-H analysis for small mass differences
(\dm\ $<30~\Gcsq$). In both cases, the dominant backgrounds 
after the cuts against the \gaga\  background (see Table~\ref{ggcuts})
are $\ww$, $\zz$ and $\ewnu$ production. These backgrounds are 
rejected
by placing a cut on the event thrust 
and on the visible mass (see 
Fig.~\ref{fig:distrib}d). The optimum cut on the visible mass can be  
parametrised with the simple form $\mvis < \dm + 5~\Gcsq$ throughout the 
full Higgsino region. 
As the AJ-H selection is applied to regions of the parameter space with large 
mass differences, the visible mass cut is less effective to reduce the 
background and cuts on the event acoplanarity are introduced. 
All selections are summarised in Table~\ref{AJcuts}.


\paragraph{Four Jets with Photons (4J-$\gamma$)}

In the mixed region, 
$\Pchipp$ and $\Pchippp$ can be produced with large cross sections. For 
large slepton masses, the cascade decays give rise to  multi-hadronic final 
states similar to 4J chargino events. The 4J-H chargino analysis is
used, with a visible mass cut re-optimised for the neutralino search. 


In this region, the radiative decay of the $\Pchip$ can be large, especially 
when 
the mass difference between the $\Pchip$ and the $\chi$ is small, giving
rise to hadronic final states with an isolated photon. A dedicated analysis 
has been designed for this topology: starting from the anti-$\gamma\gamma$ 
cuts of the 4J-H chargino analysis, an energetic isolated photon is required.
To suppress the $\zz$  and $\qqg$ backgrounds, the acoplanarity should  be 
smaller than $160^{\circ}$ and the missing momentum isolated 
($\ewedge < 7.5\pcecm $). A variable cut on the visible mass is also imposed 
depending on the signal configuration. The background is about 36 fb (20 fb) 
at 172 GeV  (161 GeV) for any cut on the visible mass, and is reduced to less
than 5 fb when the visible mass is required to be smaller than 70 GeV. For 
example, when $\Pchip\Pchipp$ production dominates and with a branching ratio 
for $\Pchip\rightarrow \chi \gamma$ of 60$\%$, the efficiency of both the 
4J-$\gamma$ analysis and the 4J-H analysis is about $15 \%$, giving a total 
efficiency of $30 \%$ with about 40 fb of background at $\rts = 172$~GeV.
The cuts are summarised in Table~\ref{AJcuts}.   
 
\paragraph{Acoplanar Leptons Topology (AL-$\chi$)}

When sleptons are light, the production of $\chi\Pchip$ or $\chi\Pchipp$ pairs 
followed by the decay 
$\Pchip\rightarrow\ell^+\ell^-\chi,\ \Pchipp\rightarrow\ell^+\ell^-\chi$ 
has a sizeable rate and leads to final states containing two acoplanar leptons 
with the same flavour and missing energy. This topology resembles the 
production of slepton pairs, therefore similar selections to those described 
in Ref.~\cite{sleptons} are used, except that the cuts against the (dominant) 
WW background are optimised for the neutralino searches as a function of the 
mass difference between the produced neutralinos. The WW background is 
reduced by requiring that the two leptons in the final state be of the same 
flavour,
and by placing a cut on the maximum momentum of both leptons. 


\paragraph{Multileptons Topology (ML)}

This analysis is used to cover the region of the parameter space  where 
sleptons are light and where also the heaviest neutralinos are produced (for 
instance, the region with  $m_0 = 75~\Gcsq$, $\Mp <130~\Gcsq$, 
$\tanb = \sqrt{2}$). In this case, several leptons can be present in the 
final state. For example, the production of $\Pchip\Pchipp$ pairs, followed by 
the decays $\Pchip\rightarrow\ell^+\ell^-\chi$, 
$\Pchipp\rightarrow\chi^{\pm}\ell\nu$, gives rise to at least three charged 
leptons in the event.  The 
following selections are designed to select these topologies with high 
efficiency. Cuts against the two-photon 
background similar to those discussed for 4J-H in Section~\ref{gghh} are 
applied, but the cut on the minimum number of charged tracks, the thrust  and 
on the maximum energy of 
the leading lepton are removed. Instead, at least two muons or two electrons 
are required (after removal of photon conversions). The leptons are ordered
in energy, where $\elepu > \elepd$.
The energies of the two leading leptons are required to be larger than 
5~GeV. Cuts on the maximum energy of these leptons and on the maximum
visible energy in the final state are also applied to reject WW and
other backgrounds.  These cuts are  optimised 
as a function of the masses of the  
neutralinos. If an isolated photon with $E>10$~GeV is present in the final 
state, as one would expect from the decay $\Pchip\rightarrow\Pchi\gamma$, then
the cut on the maximum energy of the second lepton is removed; 
this cut is directed 
against WW events which usually do not contain isolated photons.
The ML selection is summarised in Table~\ref{AJcuts}.
     



\begin{table}
\begin{center}
\begin{tabular}{|c||c|c|c|}
\hline
 \multicolumn{4}{|c|}{Neutralino} \\ 
\hline
\hline
 Topology & AJ - H & AJ - L & 4J-$\gamma$ \\
\hline
\hline
 anti-$\gamma\gamma$ cuts &  Yes  & Yes & Yes (4J-H) \\
\hline 
 $\mvis^\dagger$ & $<45$ GeV/c$^2$ & $<35$ GeV/c$^2$ & $<70$ GeV/c$^2$ \\
\hline 
 Thrust    & \multicolumn{2}{|c|}{$<0.95$} &  \\
\hline 
 $\aco$    & $<170^\circ$ &   & $<160^\circ$ \\
\hline 
 $\act$    & $<170^\circ$ &   &  \\
\hline
$\ewed$   &  &  & $<7.5\%\sqrt{s}$ \\
\hline
isolated $\gamma$   &  &  & $1$ \\
\hline
 \multicolumn{3}{c}{} \\ 
\hline
 Topology & \multicolumn{3}{|c|}{ML} \\
\hline
\hline
 anti-$\gamma\gamma$ cuts &  \multicolumn{3}{|c|}{Yes (subset of 4J-H)} \\
\hline
 identified $e/\mu$ &  \multicolumn{3}{|c|}{$\geq$2e or $\geq 2\mu$ } \\
\hline
 $\elepu^\dagger$ &  \multicolumn{3}{|c|}{$\in[5,50]$ GeV } \\
\hline
 $\elepd^\dagger$ &  
        \multicolumn{3}{|c|}{$>5$ GeV, $\in[5,25]$ GeV if $\egam<10$ GeV } \\
\hline 
 $\mvis^\dagger$ & \multicolumn{3}{|c|}{$<60$ GeV/c$^2$} \\
\hline
\end{tabular}
\end{center}
\caption{\em Cuts for neutralino analyses; cuts against $\gamma\gamma$ are 
         listed in Table~\ref{ggcuts}. The positions of the cuts for variables 
         with a $^\dagger$ depend on the point in the SUSY parameter 
         space; typical values are given for illustration 
         ($\dm=40~\Gcsq$ for AJ-H, $\dm=30~\Gcsq$ for AJ-L, 
          $\tan\beta=\sqrt{2}$, $M_2=50~\Gcsq$, $\mu=-68~\Gcsq$, 
          $m_0=75~\Gcsq$ and $\sqrt{s}=172$ GeV for ML).}
\label{AJcuts}
\end{table}

\begin{table}
\begin{center}
\begin{tabular}{|c||cc|cc||c|c|}
\hline
 Topology & \multicolumn{2}{|c|}{$\sigbg$(161)} 
          & \multicolumn{2}{|c|}{$\sigbg$(172)}  & 
  $\epsilon$  & signal($M_{\PCha}$,$M_{\Pchi,\PSnu}$) \\ 
\hline
\hline
 \multicolumn{7}{|c|}{Chargino - 4J} \\
\hline 
 VH & 21&fb & 34&fb & $22\%$ & (85,5) \\ 
  H & 29&fb & 37&fb & $54\%$ & (85,40) \\ 
  L &  4&fb & 15&fb & $33\%$ & (85,75) \\ 
 VL & 66&fb & 65&fb & $19\%$ & (85,80) \\ 
\hline
\hline
 \multicolumn{7}{|c|}{Chargino - 2JL} \\
\hline 
 VH & 25&fb & 20&fb & $35\%$ & (85,5) \\ 
  H &  4&fb &  2&fb & $61\%$ & (85,40) \\ 
  L &  9&fb &  3&fb & $47\%$ & (85,75) \\ 
 VL &  9&fb & 13&fb & $21\%$ & (85,80) \\ 
\hline
\hline
 \multicolumn{7}{|c|}{Chargino - AL} \\
\hline 
 AL-3  &     74&fb &      88&fb & $66\%$ & (85,45) \\ 
 AL-2  & $45-119$&fb &  $45-232$&fb & $65\%$ & (80,60) \\ 
 AL-VL &     80&fb &      80&fb & $19\%$ & (80,76) \\ 
\hline
\multicolumn{7}{c}{} \\ 
\hline
 \multicolumn{7}{|c|}{Neutralino} \\
\hline 
 AJ-H        & 22 &fb & 18 &fb & $38\%$ & \\ 
 AJ-L        & 35 &fb & 42 &fb & $24\%$ &  \\ 
 4J-$\gamma$ &  5 &fb &  5 &fb & $15\%$ &  \\ 
 ML          & 52 &fb & 51 &fb & $10\%$ &  \\ 
 AL-$\chi$   & 67 &fb & 56 &fb & $10\%$ &  \\ 
\hline
\end{tabular}
\end{center}
\caption{\em Efficiencies at $\rts = 172$~GeV and background for 
chargino and neutralino
 analyses. The chargino
efficiencies are determined considering only the topology for which that 
selection was optimised, for
$\mu = -500~\Gcsq$ and $\tanb = \sqrt{2}$, varying $M_1$ independently of 
$M_2$ to obtain the various ($M_{\PCha},M_{\Pchi}$) listed 
combinations. 
For the 2JL analyses, efficiencies are calculated 
for events with $\ell = e,\mu $. 
For the AL-2 analysis, the minimum and maximum background is given.
The background for the neutralino selections is 
determined  at the points given in Tables~\ref{AJcuts}. }
\label{EffBkgChar}
\end{table}

\subsection{Combination of Selections}
\label{sec:combination}



The many different selections developed for the chargino and neutralino
searches must be combined so that the analysis is
sensitive to all possible topologies for a large range of $\dm$ and 
decay branching ratios, without 
allowing excessive background from selections that contribute little
to the efficiency for a particular signal configuration.  The selections
are combined according to the $\Nbarnf$ prescription, by summing the 
signal efficiency and background expectations.  The global
analysis employs the combination of selections which, for a given choice
of the relevant parameters, minimises the average expected limit on the
production cross section.
Although there are discontinuities in the efficiency, the average expected
limit is continuous.

\subsubsection{Combination of chargino selections}

In the chargino analysis the relevant parameters are the chargino mass,
$\dm$, and the leptonic branching ratio. 
``Cross-efficiency'' between selections is not negligible as some selections are
sensitive to a topology other than that for which they were optimised;
for example, the 4J selections are efficient for the 2JL topology when the 
lepton is a tau.  There is also some overlap among the selections in the
expected background.


For the typical case in the
chargino analysis where
the dominant decay is through a virtual W, only 10$\%$
of the signal events will have the topology of acoplanar leptons, while 
44$\%$ will have a 2JL topology, and 46$\%$ will occur in the 4J topology.
The selection for acoplanar leptons (AL-3) has  irreducible background from
W pair production, and so the average expected limit is not optimal
when the acoplanar leptons selection is included.  

When
the  branching ratio
of $\PCha \rightarrow l \nu \Pchi$ is greater than 60$\%$, assuming
equal branching ratios to e, $\mu$, and $\tau$ leptons, 
a better limit is expected when the AL-3 selection is applied with 
the 4J-L, 2JL-H and 2JL-L selections. 
If stau mixing is allowed, an increase in the number of decays with $\tau$ 
leptons will be observed.  The hadronic tau decays are more efficiently 
selected by the 4J selections, so a higher branching ratio to leptons is
required before the AL-3 selection is included.  Two-body decays, $\chstau$,
can also occur when stau mixing is considered, and the same combination of 
selections is applied 
for the resulting topology.

For two-body decays to sneutrinos, $\chsnu$, the AL-2 and AL-VL selections 
are applied to optimise
the expected limit as a function of $\dm$ ($\mcha - \msnu$)  and 
$M_{\PCha}$; the AL-VL selection
is applied when $\dm < 8~\Gcsq$, for all $\mcha$.

The optimisation for charginos is performed with the
integrated
luminosity at
172~GeV, finding the optimal combination of selections for a chargino with
mass of 85~$\Gcsq$, as the chargino
limit is determined by that data set in most cases.  The optimal 
combinations
of selections for the 161~GeV data are then found for an 80~$\Gcsq$
chargino by taking the
expected 172~GeV results into account. Consequently, not all selections are 
used in the analysis of the 161~GeV data.

The combinations of selections applied to the data in the chargino analysis 
and the expected background estimates are summarised 
in
Table~\ref{table:comb-chargino}.



\subsubsection{Combination of neutralino selections}

In the neutralino analysis the optimal combination depends on the region
of the MSSM parameter space since the 
production processes and decay modes vary throughout this space.  
             
       In the Higgsino region the dominant topology  
       is acoplanar jets, irrespective of 
       $m_{0}$ and $\tanb$, and a combination of the 
       AJ-H and AJ-L analyses 
       is used for $\dm < 30~\Gcsq$.          
       For  $\dm > 30~\Gcsq$ 
       the AJ-H analysis alone is used.       
         
       In the mixed region, $\Pchipp$ and
       $\Pchippp$ production is kinematically accessible 
       and, in addition, the
$\Pchip$ has a large  radiative branching fraction when $\mu < 0$. 
       Thus for large 
       slepton masses ($m_0 \sim 200~\Gcsq$), where the neutralinos give rise 
       mainly to
       hadronic final states, a combination of the 4J-H chargino analysis
       with the 4J-$\gamma$ analysis gives the best sensitivity.
       For small slepton masses ($m_0 = 75~\Gcsq$, for example), 
       the leptonic branching ratios are enhanced and the
       AJ-H analysis is combined with the AL-$\Pchi$ and ML 
       analyses over the mixed and gaugino regions.
	Since in these regions the neutralino production
       processes and the leptonic and hadronic branching ratios
       change rapidly as a function of the parameters, the
       combination of the AJ-H, AL-$\Pchi$ and ML selections allows for a 
       robust analysis and for a stable signal efficiency. 

The combinations of selections, efficiencies, and background measurements
 for the neutralino searches are summarised in 
Table~\ref{table:comb-neutralino}.


%

\begin{table}
\begin{center}
\begin{tabular}{|c|c|c|c|}
\hline \multicolumn{4}{|c|}{Charginos} \\
\hline\hline
 \multicolumn{4}{|c|}{172 GeV}  \\ \hline
  & $\dm$ ($\Gcsq$)  & Combinations  & $\sigma_{\mathrm{bg}}$~(fb)  \\
\hline
1  & $< 10$    & 4J-VL,~2JL-VL,~2JL-L                 &  81  \\
2  & $10 - 50$ & 4J-L,~2JL-L,~4J-H,~2JL-H &  53 \\
3  & $\geq  50$ & 4J-H,~2JL-H,~4J-VH,~2JL-VH & 94  \\
\hline
4  &  all                   & 4J-L,~2JL-L,~2JL-H,~AL-3  & 101  \\
\hline \hline
\multicolumn{4}{|c|}{161 GeV} \\ \hline
  & $\dm$ ($\Gcsq$)  & Combinations  & $\sigma_{\mathrm{bg}}$~(fb)  \\
\hline
1  & $< 10$    &   2JL-VL,~4J-L,~2JL-L  & 23       \\
2  & $10 - 60$ &  4J-L,~4J-H,~2JL-H     & 34       \\
3  & $\geq 60$ & 2JL-H,~4J-VH,~2JL-VH  & 49      \\
\hline
4  &  all                   &  4J-L,~2JL-H,~AL-3     & 90       \\
\hline
\end{tabular}
\caption[.]{\em Combinations of selections used to set limits with the 
161 and 172~GeV
data, for three-body decays of charginos. For combinations listed in Rows 1-3,
 $\mathrm W^{*}$ 
branching ratios are assumed, 
and the combinations in Row 4 
are applied
when the branching ratio of $\PCha\rightarrow\ell\nu\Pchi$ is 
greater than $60\%$.  Background estimations for the combinations 
are also given.
\label{table:comb-chargino}}
\end{center}
\end{table}


\begin{table}
\begin{center}
\begin{tabular}{|l|c|c|c|c|c|c|}
\hline \multicolumn{7}{|c|}{Neutralinos} \\ \hline\hline
\multicolumn{3}{|c}{} &  \multicolumn{2}{|c|}{172 GeV} & \multicolumn{2}{|c|}{161 GeV} \\ \hline
& Region  & Combinations  & $\epsilon~(\%)$ &
 $\sigma_{\mathrm{bg}}$~(fb) & $\epsilon~(\%)$  & $\sigma_{\mathrm{bg}}$~(fb) \\
\hline
1 &  Higgsino:$\dm < 30~\Gcsq$ & AJ-L,~AJ-H    &  39 &   46  &  40  &  41  \\
2 &  Higgsino:$\dm > 30~\Gcsq$ &      AJ-H     &  38 &   18  &  41  &  22  \\
3 &  mixed:high Br$(\Pchip\rightarrow\Pchi\gamma)$ & 4J-H,~4J-$\gamma$ & 30  & 38  & 30  & 30 \\
4 &  mixed:low $m_0$ & AJ-H,~ML,~AL-$\Pchi$  & 23 & 108  & 23  & 95 \\
\hline
\end{tabular}
\caption[.]{\em Combinations of selections used to set limits with the 
161 and 172~GeV
data, for the neutralino searches.  The parameters for efficiency and
background measurements
are the same as in Table~\ref{AJcuts}.
\label{table:comb-neutralino}}
\end{center}
\end{table}


%
\subsection{Efficiency Parametrisation}
\label{sec:eff}

The efficiency of the chargino selections at its simplest level is governed
by the visible mass of the event.  This is highly correlated with 
$\dm$, which, when sleptons are heavy, is
equivalent to the maximum invariant mass $Q^2$ of the 
fermion pair from the decay
of the virtual W. 
However, the field content of the charginos and neutralinos can
affect the selection efficiency independently of $\mcha$ and $\mchi$.
 The gaugino and Higgsino components for both the $\PCha$ 
and $\Pchi$ play a role in the decay amplitude.  The CP eigenvalues 
(embedded in the neutralino mass matrix) can influence the differential
decay rate $d\Gamma/dQ^2$~\cite{chaprod}.  Aside from discrete differences
in CP eigenvalues, the $\widetilde{\mathrm W}^+$ and $\widetilde{\mathrm H}^+$ 
components in the decay amplitude depend on 
the model parameters 
in different ways,
showing up as significant changes in $d\Gamma/dQ^2$,
even for constant $\mcha,\mchi$ and dominantly gaugino-like charginos and 
neutralinos.
These effects have the largest impact on the efficiency when 
$\dm \approx 50~\Gcsq$, and can lead to differences in the efficiency of 
up to $30\%$ (relative) in some cases.

For the selections described here, the efficiencies are derived
separately for $\tanb = \sqrt{2}$ and 35.  The efficiencies are $\sim 5\%$ 
lower
 for the latter, when $\dm \sim 40-50$~GeV.  These differences are
verified with both DFGT and SUSYGEN. 

To map out the dependence of the efficiency as a function of $\dm$, $\mu$ is 
fixed to a large value, and $\tanb$ is fixed to  $\sqrt{2}$ and 35.  
After finding $\Mp$ for 
a given $\mcha$, $M_1$ is varied (thereby violating the standard
gaugino mass unification relation) to obtain the full 
range of $\mchi$ (hence, $\dm$).  Many
points in $\dm$ are generated using the full detector
simulation, with a statistical error of $\sim 1\%$ for each $\dm$ point.

The efficiencies are parametrised as functions of $\dm$ in the three
ranges of $\dm$ given in Table~\ref{table:comb-chargino}.  
They are also parametrised for the three
chargino decay topologies: ``QQ'', when both
charginos decay to $q\bar{q}'\chi$, ``QL'', when
one chargino decays to $\ell\nu\chi$ ($\ell = e,\mu,\tau$) and 
the other to $q\bar{q}'\chi$, and ``LL'', when both 
charginos decay to $\ell\nu\chi$. This is done to allow for variations
of the leptonic branching ratio, Br($\chlep$).  

Separate parametrisations are obtained for $\rts$ = 161 and 172~GeV.
A corrective factor (one for each $\dm$ range) is derived for the 
$\mcha$ dependence at fixed $\rts$.  Corrections are applied for
increases in the relative fraction of chargino decays to taus in 
proportion to all leptons.  Corrections are also applied for the 
systematic reduction in efficiency as discussed in 
Section~\ref{systematics}.  

\begin{figure}
\begin{center}
\unitlength=1mm
\begin{picture}(150,30)(0,0)
\put(17,-11){\makebox(0,0){a)}}
\put(83.5,-11){\makebox(0,0){b)}}
\unitlength=1pt
\end{picture}
\mbox{\epsfig{file=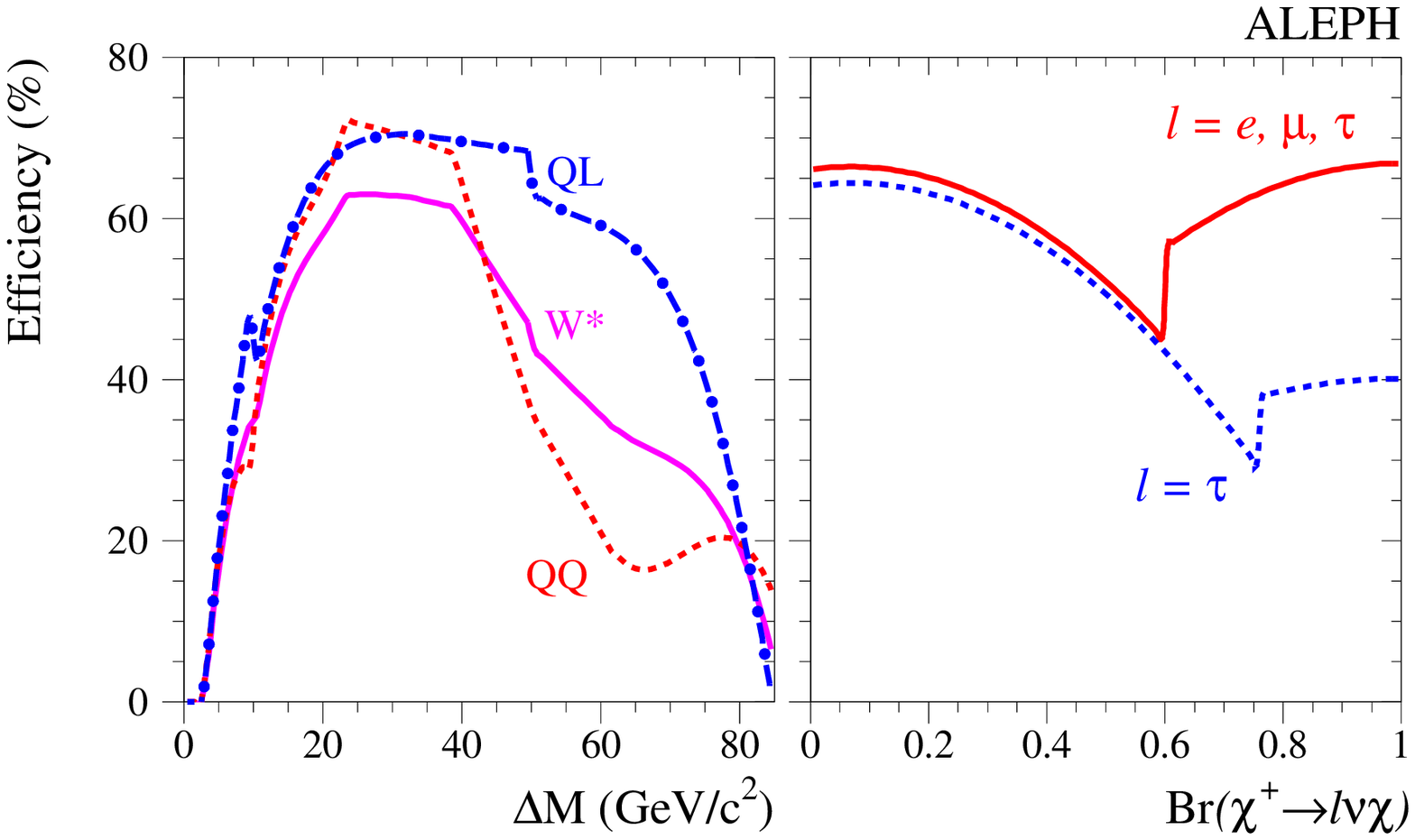,
 height=10cm%
,bbllx=5mm,bblly=20mm,bburx=190mm,bbury=140mm}}
\caption{\em Parametrised selection efficiency for the chargino 
selections listed in Table~\ref{table:comb-chargino}.  
a) The efficiency is plotted as a function of $\dm$, for  
$M_{\PCha} = 85~\Gcsq$ at $\rts = 172$~GeV, calculated 
for $\mu=-500~\Gcsq$ and 
$\tanb = \sqrt{2}$, using the combinations for W branching ratios.
Efficiencies are plotted for mixed (``QL'') and hadronic (``QQ'') topologies,
and combined assuming W branching ratios,
Br($\PCha \rightarrow \ell\nu\Pchi) = 0.33$ (``$W^*$''). 
b) The efficiency is plotted as a function of 
Br($\PCha \rightarrow \ell\nu\Pchi$), assuming equal lepton flavours 
(``$\ell = e, \mu, \tau$'') and that all leptons are taus (``$\ell = \tau$''), 
for $\dm = 40~\Gcsq$.
\label{fig:eff-cha}}
\end{center}
\end{figure}

The 
efficiencies as a function of $\dm$ and 
Br($\PCha \rightarrow \ell\nu\Pchi$)
for the chargino analysis are shown in Fig.~\ref{fig:eff-cha}.
The efficiency is shown as a function of $\dm$ for  
the combinations in rows $1-3$ of  Table~\ref{table:comb-chargino}.
Efficiencies are shown separately for the ``QQ'' and ``QL'' signal 
topologies.  The efficiency for the ``LL'' topology  
is essentially zero for these combinations of
selections. The efficiencies for the different topologies are 
combined according to the appropriate branching ratios to give an overall
efficiency; here, 
W branching ratios are applied. 
Also shown is the efficiency as a function of 
Br($\PCha \rightarrow \ell\nu\Pchi$) under the assumption of equal branching 
ratios to e,~$\mu$ and $\tau$, and assuming 100$\%$ branching ratio 
to~$\tau\nu\Pchi$, 
which gives the most conservative efficiency if stau mixing is allowed. The 
sudden increase in the efficiency at large Br($\chlep$) is 
due to the  
effect of including the AL-3 selection.

\begin{figure}
\begin{center}
\mbox{\epsfig{file=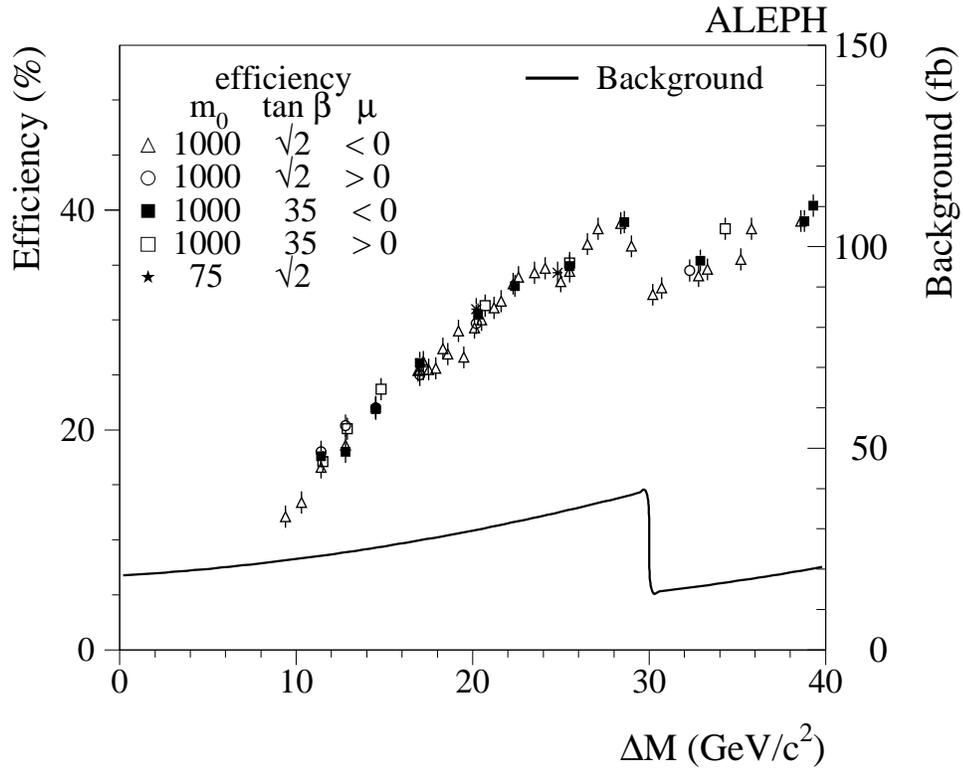,
 height=12cm%
,bbllx=1mm,bblly=1mm,bburx=180mm,bbury=180mm}}
\caption{\em  The  $\Pchi\Pchip$ selection efficiency for 
    AJ  selections in the Higgsino region 
     and expected background 
     as a function of the mass difference between the two neutralinos,
     for $\sqrt{s} = 161$~GeV. Efficiencies are similar at  
     $\sqrt{s} = 172$~GeV.
     The different symbols refer to different values of the model parameters. 
\label{fig:eff-neu} }
\end{center}
\end{figure}

The selection efficiencies for
AJ and  two-body chargino decays ($\chsnu$ and $\chstau$) are
parametrised similarly.
The efficiency for the AJ analysis is shown in 
     Fig.~\ref{fig:eff-neu} as a function of $\dm$  
     for different values of the MSSM parameters 
     ($m_{0}$, tan$\beta$, $\mu$) and for $\sqrt{s} = 161$~GeV. 
     For this analysis, in the Higgsino region,
     the efficiency depends only on the 
     mass difference between the two neutralinos, \em{i.e.},
\rm     on the visible energy in the final state, and not
     on other parameters of the theory. 
     In this region, independent of the  other model parameters, 
     the only kinematically accessible 
     process is $\Pchi\Pchip$ production and the dominant
     decay of the $\Pchip$ is $\Pchip\rightarrow \mathrm Z^{*}\Pchi$.  
     The efficiency improves at large \dm\, 
     due to the larger visible energy in the final state, and drops
     at $\dm =30~\Gcsq$ because of the change in the combination of selections.

\subsection{Studies of Systematic Effects}
{\label{systematics}}
%
%
%
%



The most important systematic effects for the 
chargino and neutralino   analyses are those which affect 
the
signal efficiency, including modelling of the signal process and detector.

The requirement that no energy be reconstructed within 
$12^{\circ}$ of the 
beam axis introduces an inefficiency due to beam-associated and detector
background 
not simulated by Monte Carlo, as it 
depends on the beam conditions during data taking.  This loss is 
measured from events triggered at random 
beam crossings to be 4.1$\%$ in the 161~GeV data and 2.4$\%$ in
the 172~GeV data.  The efficiency for the relevant selections is reduced 
accordingly.


To check the simulation of the detector response to events which 
are kinematically similar to the signal events, a sample of events
from LEP~1 is selected.  These events have an isolated 
energetic photon from final state radiation, which is removed from the
analysis of the rest of the event, leaving an acoplanar hadronic 
system with missing energy and visible mass similar to signal events.
Kinematic quantities such as thrust, transverse momentum, 
acoplanarity, isolation of the missing momentum vector, and the neutral
hadronic energy fraction are well reproduced by the Monte Carlo.  The 
simulation
of kinematic quantities for low visible mass systems is tested by comparison
to $\mathrm Z\rightarrow \tau^+\tau^-$ events, and good agreement is found.

\rm
The identification of electrons and muons has been compared in data
and 
Monte Carlo.  
The electron identification efficiency for the  2JL-VH, H, and L 
has a
systematic uncertainty of $0.6\%$ per electron. 
The selection  for the 
2JL-VL analysis 
has a systematic uncertainty associated with the electron identification 
efficiency at low momenta due to the simulation of the calorimetric and $\dedx$ estimators 
of $3.7\%$, with an additional
correction of $-3.9\%$ (relative) applied to the efficiency.
 Systematic uncertainties in the muon identification 
lead to an error of $0.7\%$ in the 2JL-VH, H, and L selections, 
$1.4\%$ in the 2JL-VL selection.

The Monte Carlo program used for simulating the chargino signal, DFGT,
has been compared to  SUSYGEN, and 
good
agreement
found for kinematic variables and signal efficiencies predicted by
the two programs.  The effects of the spin of the charginos is evident in the 
angular distribution of the leptons; however, this has an insignificant effect
on the overall efficiency.
The DFGT program 
does not include a simulation of final state radiation.  The effect of
this on the selection efficiency varies from ${\simeq 1\%}$ for high $\dm$
to 3.5$\%$ for very low $\dm$.
Signal efficiency measurements are 
corrected for this effect.

The measurement of the luminosity and beam energy can 
introduce an error in the derivation of an upper limit on the signal.
The uncertainty on the 
measurement of the integrated luminosity recorded by the
detector is less than 1$\%$ including statistical and 
systematic uncertainties.  
The beam energy is known to within  
30~MeV~\cite{lepecal}, 
causing a negligible uncertainty in the 
results of this analysis.

Systematic errors are taken into account in the derivation of 
the results for the chargino and neutralino analyses 
by means of the method detailed in 
\cite{Cousins}.  
In addition to the systematic uncertainties, statistical errors from 
the Monte Carlo statistics and the luminosity measurement, which have 
uncertainties of $\la 1\%$ each are 
taken into account.  

\section{Results}
\label{sec:results}
\subsection{Events selected in the data} 
In the $21~\pbinv$ data taken at $\rts = 161-172$~GeV,  
9.5 events are expected from background in the  
chargino selections and  5.7 in the 
neutralino selections.
There is some overlap in the background expectations for
the chargino and neutralino analyses,
  leading to a total of 13 events expected.
 A total of 15 events is observed in the data,
with some events selected by 
 both the chargino and neutralino analyses, and several events selected by
other ALEPH searches for supersymmetry~\cite{sleptons,stops}.  
A summary of
the events selected by each analysis, along with a Standard Model 
hypothesis for
each candidate, is given in Table~\ref{table:cha_cand}.
The numbers of events expected and observed by the various combinations of 
selections is given in Table~\ref{table:Nexp}.

In the chargino analysis, a total of 
3.7 events are expected to be selected by the 4J and 2JL analyses in the
161 and 172~GeV data set, and 
five events are observed.
Two events are selected by the 4J-VL analysis in the 172~GeV 
data set; both are consistent with $\gamma\gamma$ background.  
One of these events has missing momentum pointing to the vertical
LCAL crack; an undetected electron from a tagged $\gaga$
event is a possible explanation.  The other event has an energy deposit in 
the HCAL 
which, possibly due to incorrect reconstruction, is not associated to a 
track, giving the event 
``extra'' 
transverse momentum.  Two events are selected by the 2JL-VH and 4J-VH 
selections, 
one each in the 161 and 172~GeV data; both are compatible with 
WW production.  One
event can be interpreted as $\ww\rightarrow\tau\nu q\overline{q}$, 
where
the tau decays hadronically.  Due to a nuclear interaction in the ITC/TPC 
wall, 
the kinematics of the event are mismeasured.  The other event can also
be interpreted as
$\ww\rightarrow\tau\nu q\overline{q}$, where the tau decays to
an electron
and neutrinos.  A possible mismeasurement of the energy of a 
low-angle jet 
due to 
cracks in the detector allows this event to be selected.  The 
kinematics of the 
event 
selected 
by the 4J-L analysis in the 161~GeV data 
suggest its origin 
as the four-fermion process 
$\ZZg\rightarrow\Pnu\Panu\tau^{+}\tau^{-}$ 
where the taus decay to $\rho$ and $a_{1}$.  This event is
also selected by the searches for neutralinos, stops, and staus.

In the AL-3 selection, three events are 
observed in the 172~GeV data, 
while
1.6 are expected in the entire data set.  One of these events is also selected
by the smuon search, and is compatible 
with $\ww$ or 
$\zz$ production.  The other two events are consistent with 
$\ww\rightarrow\tau\nu \tau\nu $, with one-prong tau decays.
The events are not selected by the slepton searches because the  
tracks are not both identified as leptons.
In the AL-2 selection, 
the same three events are selected by the high mass difference analysis, 
while 3.8 events are expected from the 161 and 172~GeV data set.
Five events, compatible with background processes, are selected by the 
AL-VL analysis, while
two are expected, as described 
in Ref.~\cite{sleptons}.


        In the neutralino analyses, two events are selected by the 
    AJ-L and AJ-H analyses, 
   while 1.9 are expected. Both events are 
     among those selected by the chargino analysis. 
One event is observed in the data by the combination 
     of the 4J-H and the 4J-$\gamma$ analyses, while 0.8 are expected.
     This candidate is selected by the 4J-H analysis only for the neutralino 
     case, where the optimisation procedure leads to a visible mass cut above
     $70~\Gcsq$, in contrast to the chargino search. This event has a 
     visible 
     mass of $70.3~\Gcsq$ and shows a large charged track multiplicity. There is a 
     clear sign of an isolated minimum ionising particle at low angle, with 
     a muon-like digital pattern in the HCAL and one hit in each muon chamber 
     layer. No charged track is reconstructed because the particle is at a very
     low angle: only two TPC hits are recorded. This favours the $\ww$ 
     interpretation of this event, with a leptonic decay of one W. 

     Finally, two events are 
     selected while three are expected in  the 
     combination of the AJ-H, AL and ML analyses optimised 
        for the neutralino search for small $m_0$ in the mixed region.
One of them  
     is common to the chargino candidate sample. The other one is a 
     slepton candidate and is described in detail in 
     Ref.~\cite{sleptons}.

\begin{table}
\begin{center}
\begin{tabular}{|l|c|c|l|}\hline
 &\multicolumn{2}{|c|}{Selection} & \\ \hline
$\rts$~(GeV) & Chargino  & Neutralino  & Hypothesis \\ \hline
        161  & 4J-VH     &             &     $\ww\ $                              \\
             & AL-VL     &             &    $\gagall$                    \\
             &           & AL-$\Pchi$,ML &     $\ww,~\zz$                   \\
             & 4J-L      & AJ-L        &     $\ZZg  \rightarrow \tau\tau $   \\
\hline
 172         & 2JL-VH    &             &     $\ww$                               \\
             & AL-VL     &             &     $\gagall$                      \\
             & AL-3,AL-2 &    ML       &     $\ww,~\zz$                      \\
             &           &  4J-H       &     $\ww$                         \\
             & 4J-VL     &             &     $\ggqq$                      \\
             & AL-VL     &             &     $\gagall$               \\
             & AL-3,AL-2 &             &     $\ww$                               \\
             & AL-3,AL-2 &             &     $\ww$                             \\
             & 4J-VL     &    AJ-L     &     $\gaga\rightarrow\tau\tau$     \\
             & AL-VL     &             &     $\gagall$                      \\
             & AL-VL     &             &     $\gagall$                      \\
\hline

\end{tabular}
\caption[.]{\em Candidate events selected by the chargino and neutralino 
analyses in the 161 and 172~GeV data.
\label{table:cha_cand}}
\end{center}
\end{table}

\begin{table}
\begin{center}
\begin{tabular}{|r|c|c||c|c|c|}\hline
 \multicolumn{3}{|c||}{Chargino} &\multicolumn{3}{|c|}{Neutralino} \\ \hline
\hline
Combination & ${\mathrm N_{exp}}$ & ${\mathrm N_{obs}}$ 
& Combination & ${\mathrm N_{exp}}$ & ${\mathrm N_{obs}}$ 
\\ \hline
 1         & 1.1  & 3           & Higgsino:~$1-2$  & 1.9   &   2   \\
 $W^*$~~2  & 0.9  & 1           & mixed:~3         & 0.8   &   1   \\
 3         & 1.5  & 2           & mixed:~4         & 3.0   &   2   \\
\hline
$\PCha(\ell)$ & 5.8  & 8      & \multicolumn{3}{|c|}{ }   \\ 
\hline
total      & 9.5  &  13         &  total           & 5.7   &   5   \\
\hline 
\end{tabular}
\caption[.]{\em Numbers of background events expected and observed
by the chargino and neutralino 
analyses in the 161 and 172~GeV data.
 In the chargino column, the ``~$W^*$'' combinations correspond to rows $1-3$ of
Table~\ref{table:comb-chargino}, and the neutralino combinations
correspond to those given in Table~\ref{table:comb-neutralino}.
``$\PCha(\ell)$'' refers to the total number of events from AL-3,AL-2, and 
AL-VL selections (see Table~\ref{EffBkgChar}).
\label{table:Nexp}}
\end{center}
\end{table}

\rm

The number of events selected in the data, their
distribution among the selections and their properties do not 
suggest a signal 
for supersymmetry.  Therefore, limits are set on the production of charginos 
and neutralinos, and constraints placed on the parameters of the MSSM.
The 
candidate events are taken into account in deriving the limits in the 
regions of 
($M_{\PCha},M_{\Pchi}$) and
($M_{\Pchip},M_{\Pchi}$) in which the analyses that 
select each candidate are applied.  
 For the combinations using the AL-2 selection only, the WW background is
subtracted~\cite{pdg} from the AL-2 selection.

\subsection{Limits on the production cross section}

\begin{figure}
\begin{center}
\mbox{\epsfig{file=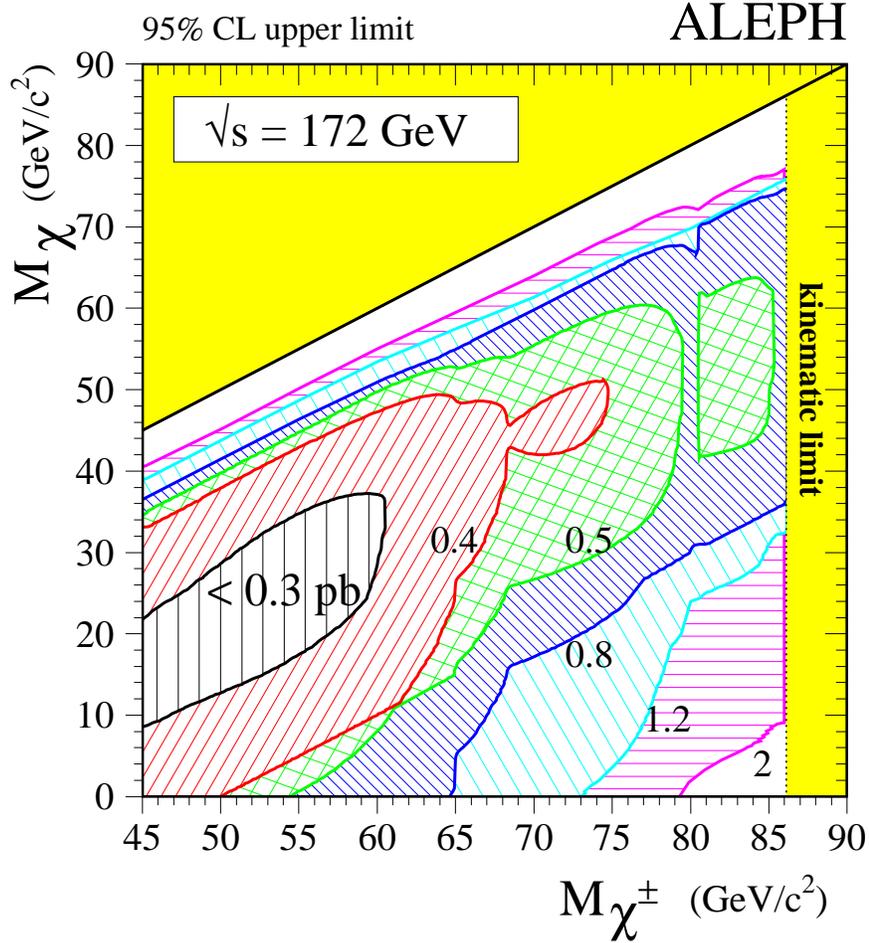,height=15cm%
,bbllx=5mm,bblly=50mm,bburx=175mm,bbury=250mm}}
\caption{\em The 95$\%$ C.L. upper limit on the cross section 
for chargino pair production, in the ($\mcha,\mchi$) plane.  Data 
taken at lower centre-of-mass energies
(130, 136, 161, and 170~GeV) are included by scaling the luminosity by the
ratio of the cross section at that energy to the cross section at 
$\rts = 172$~GeV , for $\mu = -500~\Gcsq$ and $\tanb = \sqrt{2}$. 
W branching ratios are assumed in the chargino decay.
\label{fig:chacross}}
\end{center}
\end{figure}

Upper 
limits on
sparticle production cross sections can be derived from 
the results of these searches.
Unless sleptons are light, $\mathrm W^{*}$ exchange dominates the decay  
of charginos, so the process
$\epem\rightarrow\chi^{+}\chi^{-}\rightarrow\mathrm
W^{*}\Pchi\mathrm W^{*}\Pchi$ 
defines the signal topology
used to set upper limits on the cross section  in the plane of
$\mcha$ and 
$\mchi$, shown in Fig.~\ref{fig:chacross}.
The efficiencies used in the derivation of this limit are calculated for
$\mu=-500~\Gcsq, \tanb = \sqrt{2}$, using the techniques described in 
Section~\ref{sec:eff}.
The features of the contours of constant cross section reflect
discontinuities in the number of candidates at points where the 
combinations of selections change and where the additional 
luminosity
from data taken at lower energies applies.  
The integrated luminosities taken at 
centre-of-mass energies of 130, 136~\cite{bib:paper133}, 161, and 170~GeV are 
scaled by the ratio of cross sections in the gaugino region 
($\mu = -500~\Gcsq, \tanb = \sqrt{2}$)
to those at 
172~GeV, and
included with the data taken at 172~GeV to derive this limit.
The ratio of cross sections is slightly larger in the gaugino region than
in the Higgsino region; however, the result differs by less than $10\%$.

\begin{figure}
\begin{center}
\mbox{\epsfig{file=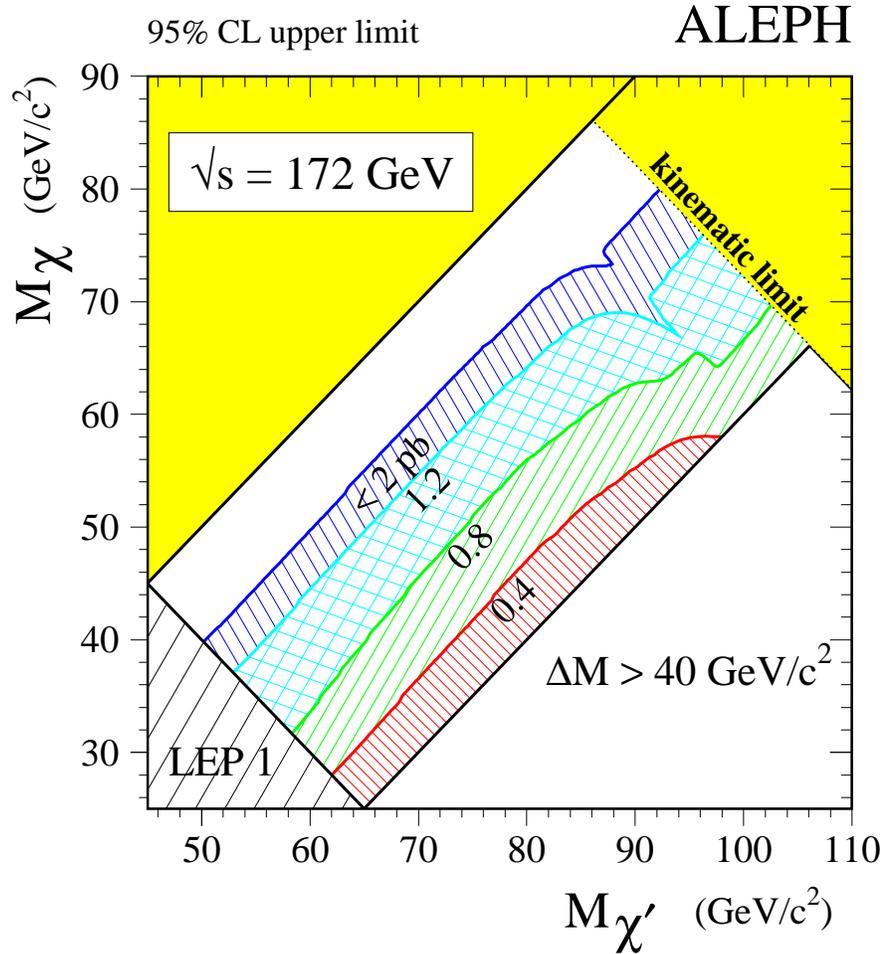,height=15cm%
,bbllx=5mm,bblly=50mm,bburx=175mm,bbury=250mm}}
\caption{\em The 95$\%$ C.L. upper limit on the cross section 
for $\Pchip\Pchi$ production, in the ($\mchip,\mchi$) plane, for the 
neutralino mass ranges relevant in the Higgsino region ($\dm < 40~\Gcsq$).  
The region where $\dm > 40~\Gcsq$ has not been investigated.
Data 
taken at lower centre-of-mass energies
(130, 136, 161, and 170~GeV) are included by scaling the luminosity by the
ratio of the cross section at that energy to the cross section at 
$\rts$=172~GeV.
Z branching ratios are assumed in the $\Pchip$ decay.
\label{fig:neucross}}
\end{center}
\end{figure}

Similarly, the neutralino AJ-H and AJ-L searches can be used to derive an 
upper limit on the cross section for $\Pchi\Pchip$ production, where the decay
$\Pchip \rightarrow \mathrm{Z}^{*}\Pchi$ is assumed. The resulting cross section
limit for the range of ($M_{\Pchip},M_{\Pchi}$) relevant in the Higgsino region
is shown in Fig.~\ref{fig:neucross}.  Values of $\dm > 40~\Gcsq$ have not 
been considered because for the given luminosity, the cross sections are too 
low to allow this region to be useful as a constraint.

\subsection{Interpretation in the MSSM}

The constraints that the results of the chargino and neutralino searches
can place on the parameters of the MSSM are explored in this section.  
First, limits are
derived assuming that sleptons are heavy. In this case, charginos and 
neutralinos decay with W and Z branching ratios, respectively.
Next, sleptons are allowed to be light, and the 
resulting changes to the cross section and decay branching ratios are 
explored.  Stau mixing can potentially affect the limits derived from the 
searches, as the decays of charginos and neutralinos are modified by the 
presence of a light stau.  These
effects have been investigated, and 
the resulting limits are only slightly modified
from previous cases.  Assumptions commonly made in SUSY GUT's are then
relaxed, and exclusion limits independent of requirements of a universal 
slepton mass are derived.  The assumption of a universal gaugino mass is 
also relaxed, and finally, limits without assumptions on a universal scalar
or a universal gaugino mass are given.

Limits are derived using parametrisations for the efficiencies, as described
in Section~\ref{sec:eff}, and the slight variations in efficiency due to 
field content are checked with the full MSSM simulation.

\subsubsection{Standard scenario: heavy sleptons}
\label{gut-hi_m0}

\begin{figure}
\begin{center}
\unitlength=1mm
\begin{picture}(150,10)(0,0)
\put(9,77){\makebox(0,0){a)}}
\put(91.5,77){\makebox(0,0){b)}}
\end{picture}
\mbox{\epsfig{file=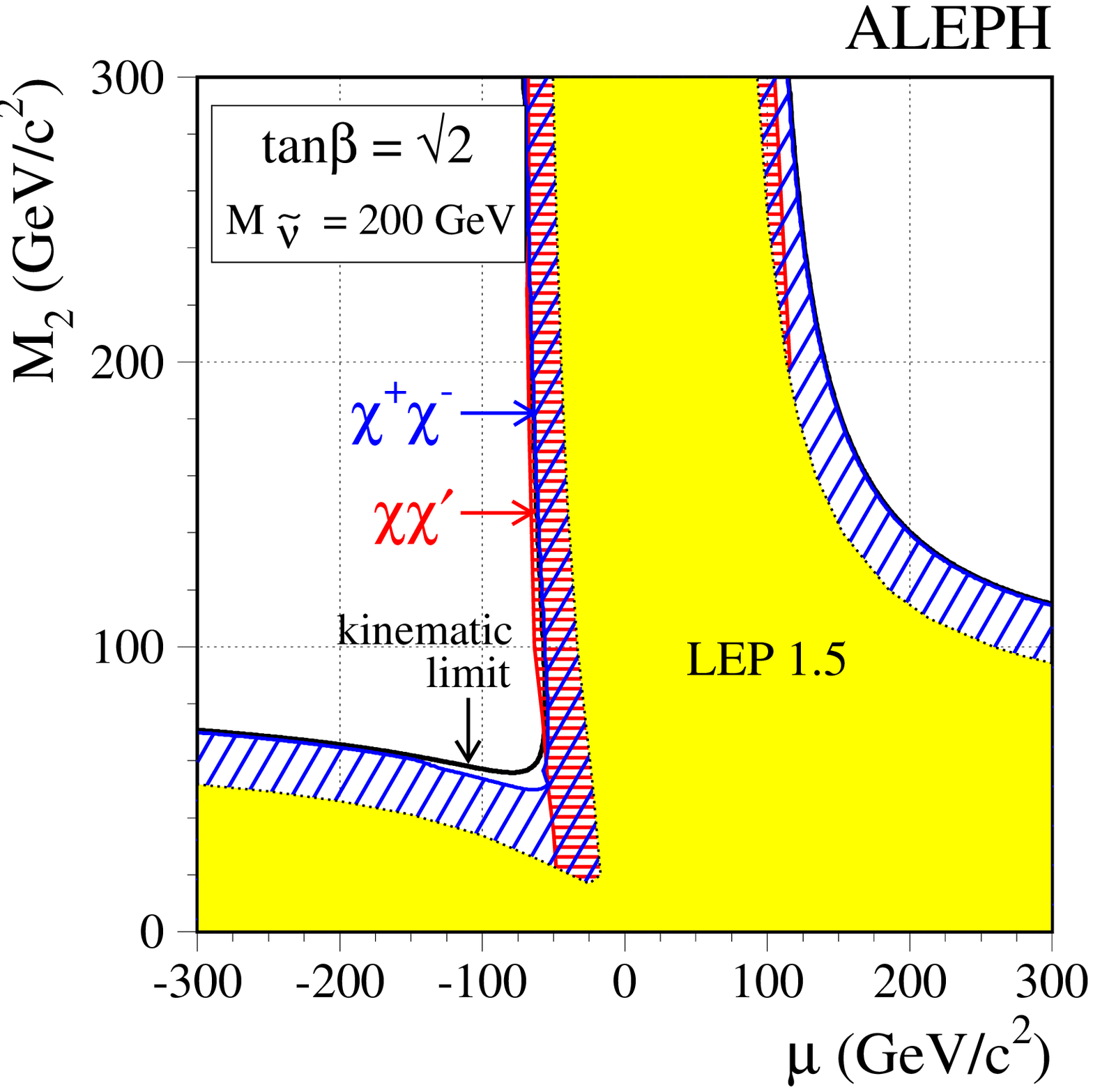,height=9cm 
,bbllx=350mm,bblly=50mm,bburx=350mm,bbury=250mm}}
\mbox{\epsfig{file=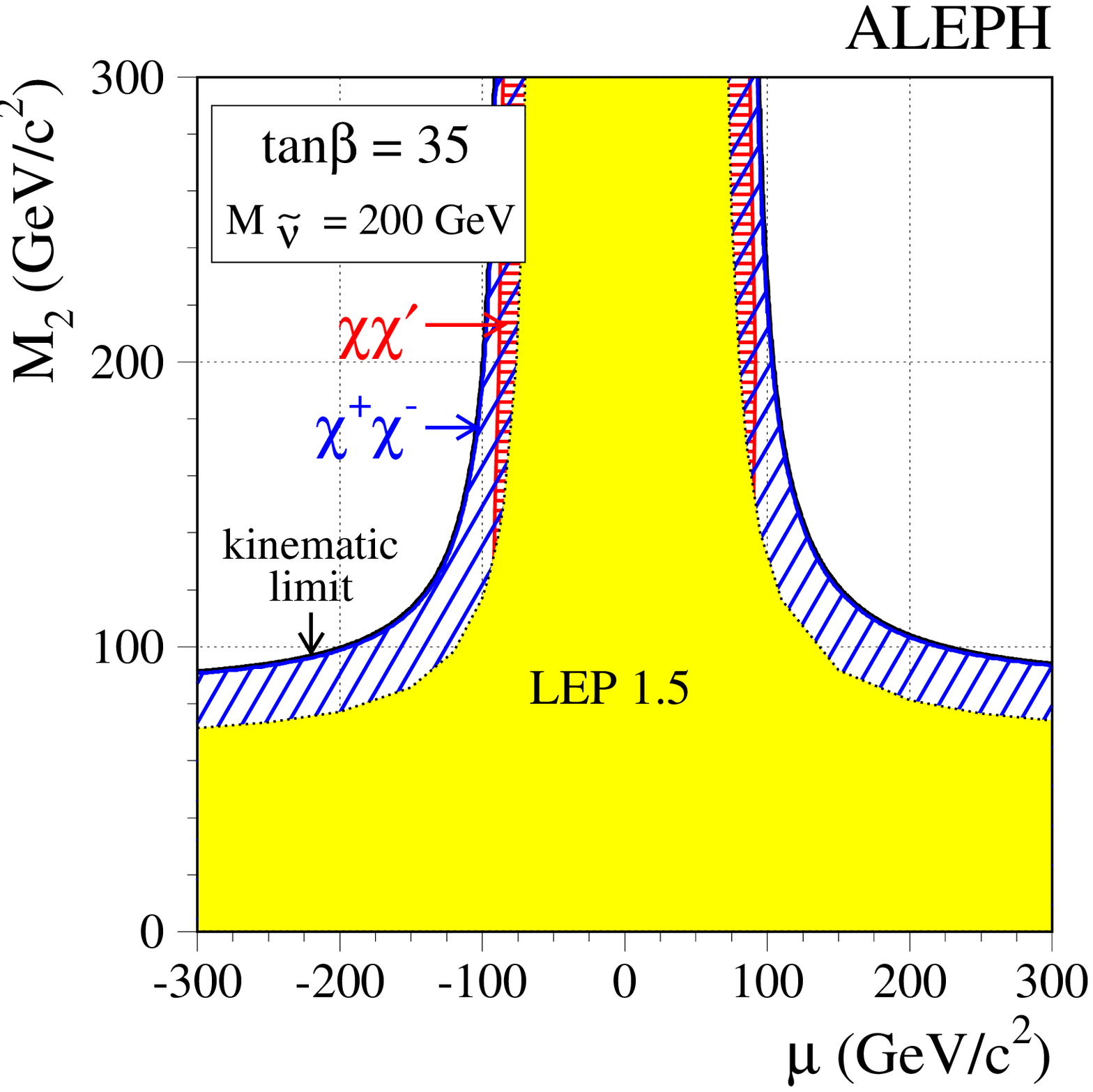,height=9cm 
,bbllx=170mm,bblly=50mm,bburx=170mm,bbury=250mm}}
\caption{\em Excluded region in the ($\mu,M_{2}$) plane, for 
a) $\tanb = \sqrt{2}$ and b) $\tanb = 35$, and $M_{\PSnu}=200~\Gcsq$.
The light grey region was obtained at LEP 1.5~\protect\cite{bib:paper133}.  
The dark line shows the kinematic limit for chargino production at 
$\rts = 172.3$~GeV.  The slanted hatched region shows the region excluded by 
the chargino search, and the horizontal hatched region, by the neutralino 
search.
\label{fig:mu-M2-tb14}}
\end{center}
\end{figure}

Chargino and neutralino masses and cross
sections
are determined by the parameters $\mu$ and $\Mp$, for 
given values of $\tanb$ and $m_{0}$.
Limits on
the production of charginos and neutralinos constrain these parameters, 
as depicted in Fig.~\ref{fig:mu-M2-tb14}
for the given values of
$\tanb$ and for $\msnu =200~\Gcsq$. 
At this value of $\msnu$, the decay branching ratios are unaffected, but 
the cross section is reduced with respect to its asymptotic
value. 
(When sleptons are heavy, detailed 
assumptions made on the relations among their masses are unimportant.)

In the gaugino region, the chargino production cross section is high, and
selection efficiency is high since  $\dm \simeq M_{\PCha}/2$ 
(see Fig.~\ref{fig:eff-cha}).  
As a result, charginos are excluded nearly to the kinematic
limit.
The limit on the chargino mass is  $85.5~\Gcsq$ for $\mu = -500~\Gcsq$ and
$\tanb = \sqrt{2}$.
 In the 
Higgsino region, the cross section is lower and $\dm$ is small, leading
to a lower selection efficiency due to the difficulties of 
rejecting $\gaga$ background, 
and a slightly weaker limit ($\mcha > 85~\Gcsq$ for $\dm > 10~\Gcsq$, 
corresponding to $\Mp \la 550~\Gcsq$).  
The additional 
gain from the
search for $\Pchi \Pchip$ production, which is most powerful in the 
Higgsino  and mixed regions for low \tanb, is also  shown. 
For high $\tanb$, 
the result from the chargino searches is similar, and 
the exclusion 
reaches the kinematic limit;
no additional
exclusion is gained from the neutralino search.  In the following, the 
discussion will be concentrated on the 
low $\tanb$ case.

\begin{figure}
\begin{center}
\mbox{\epsfig{file=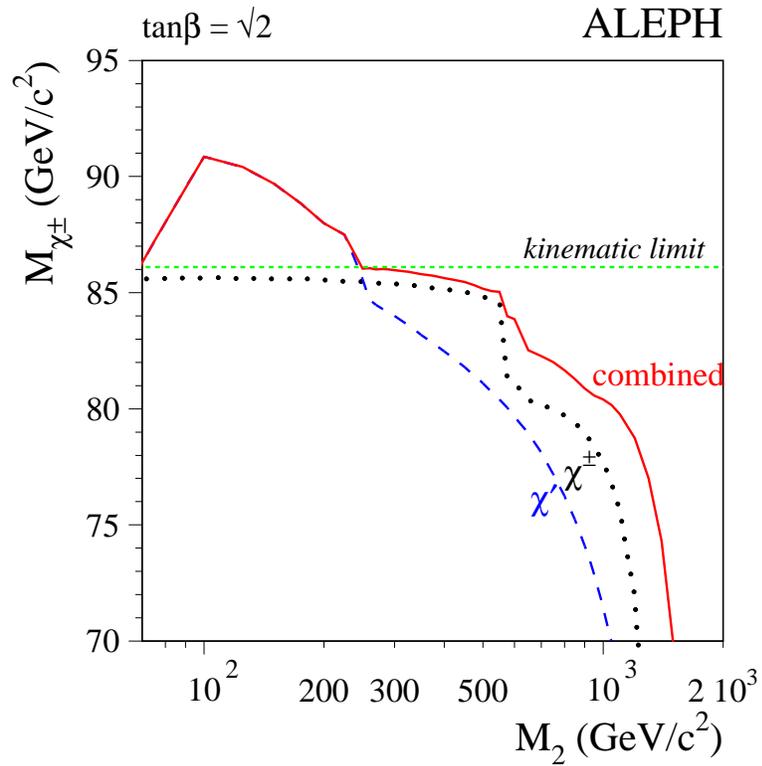,height=11cm%
,bbllx=2mm,bblly=45mm,bburx=170mm,bbury=230mm}}
\caption{\em The limit on the chargino mass as a function of
$\Mp$, from the chargino search (labelled $\PCha$), 
from the neutralino search (labelled $\Pchip$), and from the
combination of chargino and neutralino searches. This limit is derived 
for $\tanb = \sqrt{2}$ and $m_0 = 200~\Gcsq$.
\label{fig:mcha-M2}}
\end{center}
\end{figure}

The impact of the neutralino search is seen more clearly in 
Fig.~\ref{fig:mcha-M2}.  The limit on the chargino mass as a function of 
$\Mp$ is derived using the chargino and neutralino analyses separately.
 For lower $\Mp$, charginos are excluded nearly to the kinematic
limit by the chargino search alone.  The neutralino analysis allows
exclusion  beyond the kinematic limit for chargino production.  
For higher $\Mp$, $\dm$ becomes small, leading 
to a lower selection efficiency.  The abrupt reduction in the limit 
from the 
chargino search  at $\Mp \sim 550~\Gcsq$
is due to the increase in the number of candidates to be taken into account for
$\dm < 10~\Gcsq$.

Charginos and neutralinos constitute independent signals in the 
Higgsino region.
The selection criteria developed for the chargino signal do not augment
significantly the neutralino acceptance, and vice versa.  To obtain
a combined limit, expected signals and numbers of candidates are summed,
extending the exclusion in the Higgsino region, as shown in 
Fig.~\ref{fig:mcha-M2}.
This is most evident in the deep Higgsino region, where the combination of 
chargino and neutralino analyses sets a limit on the chargino
mass above 79$~\Gcsq\ $, for $\Mp \leq 1200~\Gcsq$ (corresponding to 
$\dm \geq 5~\Gcsq$).  This improves the limit of $72~\Gcsq$ set by the chargino
search alone.  

At the ``Supersymmetric Limit'', where $\tanb = 1$ and $\Mp = \mu = 0$, both 
$\chi^{\pm}$ and $\chi^{\pm}_{2}$ have mass $\simeq M_{\mathrm W}$.  The 
two lightest neutralinos are nearly massless, and $\Pchip$ decays to 
$\Pchi\gamma$ with $100\%$ branching ratio.  Production of heavier neutralinos
is also kinematically possible, with 
$M_{\Pchipp} \simeq M_{\Pchippp} \simeq M_{\mathrm Z}$.  This process was
accessible at $\rts = 130-136$~GeV, and was used to exclude the 
Supersymmetric Limit~\cite{bib:lsp-133}.
At $\rts > 2M_{\mathrm W}$, direct exclusion of this region 
using chargino searches also is possible.
Since the mass difference between
the charginos and neutralinos is $\simeq 80~\Gcsq$, the search is
difficult due to the WW background, but the efficiency of the 2JL-VH and
4J-VH selections allows 
this point to be excluded. The optimal limit is expected when only
the data taken at $\rts = 172$~GeV is included.  No background is subtracted
in the derivation of this limit.
The upper limit on the cross section is 2.9~pb, and
at $\rts = 172$~GeV, the theoretical cross section is $> 2.9$~pb in this region.
This limit is derived for $m_0 = 200~\Gcsq$.

\subsubsection{Effects of light sleptons}

The effect of low slepton masses is significant in both the 
production and decay of charginos and neutralinos, as explained in 
Section~\ref{sec:intro}.  
Here limits are derived from the chargino 
and neutralino searches when sleptons are light and the particular role played 
by staus is clarified.  Direct searches for sleptons can also play a role
in the chargino and neutralino limits in this scenario.
A general exclusion will be treated fully in a forthcoming publication;
for preliminary results, see Ref.~\cite{interpretations}.

~~\\
\bf {Light sleptons, nominal stau mixing} \rm
~~\\

\begin{figure}
\begin{center}
\unitlength=1mm
\begin{picture}(150,10)(0,0)
\put(-2.5,99){\makebox(0,0){a)}}
\put(90.,99){\makebox(0,0){b)}}
\end{picture}
\mbox{\epsfig{file=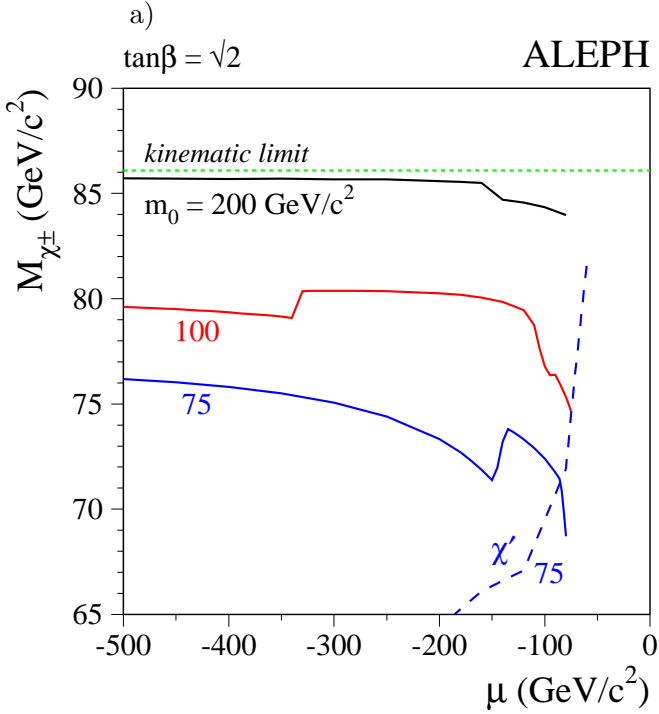,height=9cm%
,bbllx=320mm,bblly=39mm,bburx=320mm,bbury=206mm}}
\mbox{\epsfig{file=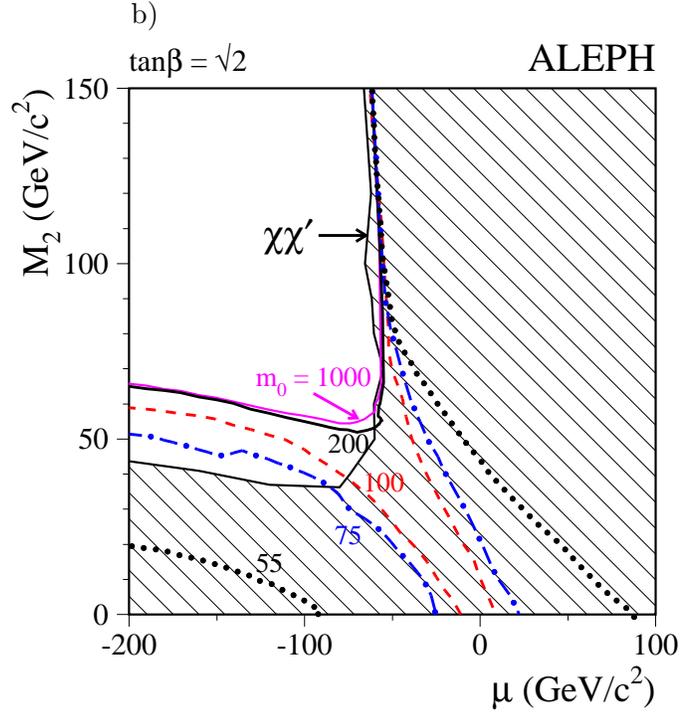,height=9cm%
,bbllx=150mm,bblly=39mm,bburx=150mm,bbury=206mm}}
\caption{\em 
a) The limit on the chargino mass as a function 
of $\mu$ for $\tanb = \sqrt{2}$, for several values of $m_0$. 
The excluded chargino mass
from the neutralino search for $m_0 = 75~\Gcsq$ is shown as a dashed curve.
The chargino exclusion is 
independent of $\mu$ when $\mu \ga -60~\Gcsq$, as can be
seen in b), where
the exclusion in the ($\mu,\Mp$) plane 
for several
values of $m_0$ is shown.  The dark curves indicate limits from the chargino
searches, and the hatched area is the exclusion from the neutralino
analysis, for $m_0 = 75~\Gcsq$. These limits are derived for 
$\tanb = \sqrt{2}$.
\label{fig:low_m0}}
\end{center}
\end{figure}

The limit on the chargino mass  throughout the gaugino 
region is
evaluated as a function of $\mu$
for several values of $\mzero $ and for $\tanb = \sqrt{2}$, as 
shown in Fig.~\ref{fig:low_m0}a.  This limit is derived 
from the chargino analysis assuming 
a universal scalar mass $\mzero$ for the sleptons.  
The overall 
reduction in the limit for decreasing $\mzero$ is due to the diminished
cross section.   As charginos become more gaugino-like (\em{i.e.}, \rm as
$|\mu|$ increases), the leptonic
branching ratio increases, and the selection for acoplanar leptons is applied to 
retain efficiency;
the sharp change in the
limit is due to the change in efficiency and number of candidates.
Stau mixing, discussed further below, is calculated with 
$\atau = 0~\Gcsq$ (at the electroweak scale).

The evaluation of low $\mzero$ effects is extended to the 
($\mu,\Mp $) plane, as shown in Fig.~\ref{fig:low_m0}b.  
The reduction of the limit
 in the gaugino region, as seen in the previous plot, is also 
evident in the  
mixed region, where a ``valley'' opens up for $-\mu \approx \Mp$.
There is a modest improvement in the excluded region 
from charginos as $\mzero$  increases from  $200~\Gcsq$ to  $1000~\Gcsq$.
  In the Higgsino region,
light scalars have little effect on the exclusion obtained with the chargino 
and 
neutralino searches, and all limits are similar to the high $\mzero$ results.

In contrast to chargino production, the neutralino cross section increases 
significantly as $\mzero$ is reduced below $100~\Gcsq$.  The enhancement of 
the leptonic branching ratios
 motivates
      the combination of the acoplanar jet analysis with 
      the multi-leptonic analyses. The results  cover the chargino valley 
  and slightly improve the 
      LEP~1 limit~\cite{opal-lep1} (not shown). In the gaugino region 
      where only $\Pchi \Pchip$ and $\Pchip \Pchip$ are produced, the 
      larger cross section for
$m_0 = 75~\Gcsq$ 
      allows the derivation of limits which are almost as constraining as the 
      chargino limits.  The limits from neutralino searches for 
$m_0 = 75~\Gcsq$ are expressed as a limit on the chargino mass in 
Fig.~\ref{fig:low_m0}a.

In chargino production and decay for low $\mzero$, the relevant physical 
quantity is the mass of the sneutrino, as this determines the reduction 
in cross section and enhancement of the leptonic branching ratio.  Therefore,
the limit on the chargino mass can be meaningfully expressed as a function
of the
sneutrino mass.  

\begin{figure}
\begin{center}
\mbox{\epsfig{file=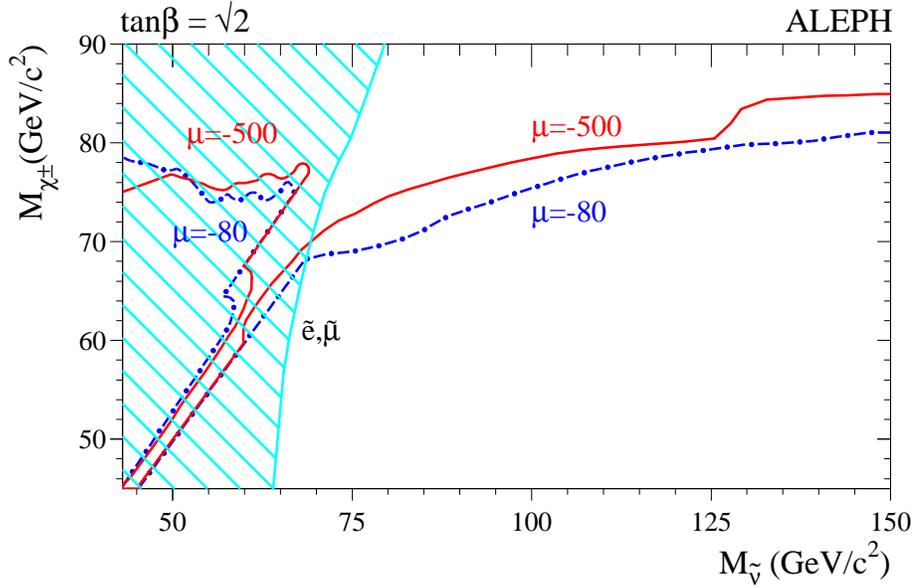,height=8cm%
,bbllx=1mm,bblly=1mm,bburx=195mm,bbury=125mm}}
\caption{\em The limit on the chargino mass as a 
function of sneutrino
mass in the gaugino region, 
for $\tanb = \sqrt{2}$ and $\atau = 0$.  
 The limit from selectron and smuon searches for
$\tanb = \sqrt{2}$ and $\mu = -80~\Gcsq$ is also indicated.
\label{fig:stau_3bdy}}
\end{center}
\end{figure}

As seen in Fig.~\ref{fig:stau_3bdy} for two points in the gaugino region, 
when $\msnu \geq 150~\Gcsq $,
there is little effect due to the sneutrino mass; this
is similar to the case under which the limits in 
Fig.~\ref{fig:mu-M2-tb14} 
are derived.  The cross section decreases as 
$\msnu$ decreases, and  the leptonic branching ratio increases, 
necessitating a change in selections to include the AL-3 selection.  
In the near gaugino region 
($\mu=-80~\Gcsq$), light sleptons have less of an effect, and the leptonic
branching ratio does not increase above $50\%$.  The limit in this case is 
lower than for $\mu=-500~\Gcsq$ due to the lower cross section. 

When $\msnu \leq \mcha$, two-body decays $\chsnu$ tend
to dominate.  
The AL-2 and AL-VL selections 
are used to derive the limit in this region.
When 
the mass difference is too small,
the leptons do not have enough
energy to pass the selection, and no exclusion is obtained.  This is the 
``corridor''
visible at low $\msnu$ in Fig.~\ref{fig:stau_3bdy}, which extends
to the LEP 1 limit of $45~\Gcsq$ on the chargino mass.  
The differences in the
corridor for these two values of $\mu$ 
are due to the effects of stau mixing, 
discussed in the following subsection.  
The features of the contour reflect the 
accounting for
candidates as a function of $\dm$.  
Also shown is the limit from the slepton search~\cite{sleptons}, which
excludes the corridor where no limit can be obtained from the chargino search.
The slepton limit is derived for 
$\tanb = \sqrt{2}$ and $\mu = -80~\Gcsq$, and is much weaker for high $\tanb$; 
thus, a general exclusion of this region is difficult.

~~\\
\bf {Effects of stau mixing} \rm
~~\\

Due to the relatively high mass of the tau lepton, mixing between the left- 
and right-handed  staus can occur, modulated by the off-diagonal term in the 
stau mass matrix, $- M_{\tau}(\atau + \mu \tanb)$.  The lightest stau, \Pstaup, 
can be significantly lighter than the other sleptons and sneutrinos, causing 
an increase in the branching ratios of charginos and neutralinos to final 
states with taus.  The decay amplitude also depends on the field content of 
the chargino or neutralino, and is most enhanced in the gaugino region.  
Thus, the  effects of stau mixing are most evident for gaugino-like 
charginos and neutralinos.  

\begin{figure}
\begin{center}
\mbox{\epsfig{file=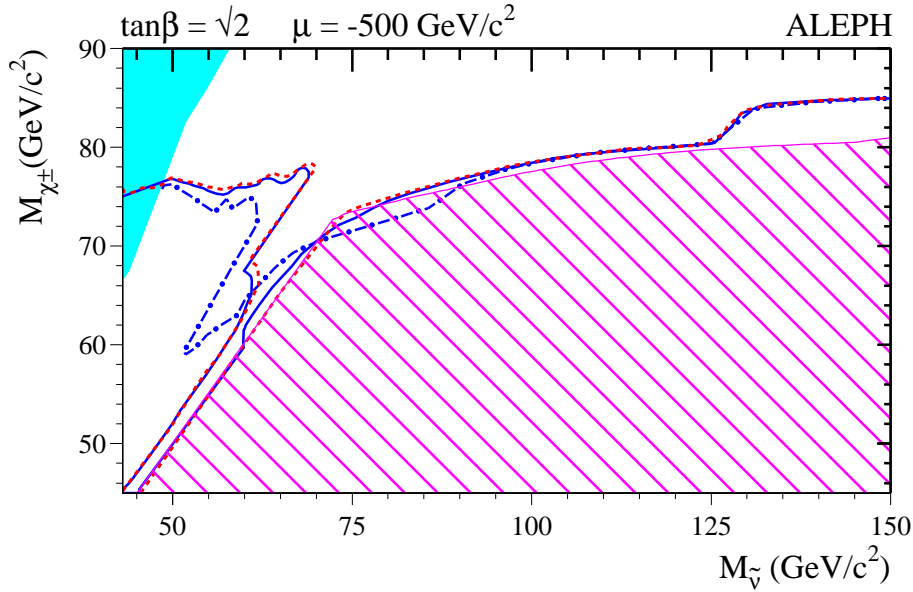,height=8cm%
,bbllx=1mm,bblly=1mm,bburx=195mm,bbury=125mm}}
\caption{\em The limit on the chargino mass as a function
of sneutrino mass, for $\mu = -500~\Gcsq$, $\tanb = \sqrt{2}$,
and various $\atau$.  
First, a universal scalar mass is
assumed, and the limit is derived for $\atau = 0$ (solid curve),
$\atau=+1~\Tcsq$ (dashed curve) and $\atau = -1~\Tcsq$ (dot-dash curve).
Second, the assumption of a universal $\mzero$ is dropped, and the impact
of stau mixing is maximised (for three-body chargino decays), shown as the
hatched region.
The shaded region at low $\msnu$ is
theoretically forbidden for $\atau = -1~\Tcsq$.
\label{fig:stau_3_A}}
\end{center}
\end{figure}

To study the effects of stau mixing, a point in the deep gaugino region is 
chosen, specifically $\mu = -500~\Gcsq$, for $\tanb = \sqrt{2}$.  A low 
value of $\tanb$ is chosen since for high $\tanb$ in the gaugino region, 
the stau mass can become unphysical ($M_{\Pstaup}^{2} < 0$), for low $\mzero$.  
The limit on the chargino mass as a function of sneutrino mass 
(still assuming a universal slepton mass) is shown in 
Fig.~\ref{fig:stau_3_A} for three values of the tri-linear coupling term, 
$\atau$:  $\atau = 0$ and $\atau = \pm 1~\Tcsq$.

For high $\msnu$, there is little effect from stau 
mixing, and the results are as discussed previously.  
As $\msnu$ decreases, the slepton masses also decrease, with a comparatively
larger impact of the mass splitting in the stau sector.  However, even if the
detailed behaviour of the chargino mass limit as a function of $\msnu$ is
affected by the precise value of $\atau$, the global features observed for
$\atau = 0$ remain.  The differences shown in Fig.~\ref{fig:stau_3_A} are due to the 
mass of $\Pstaup$ and its coupling to the chargino.  Decays of $\chstau$ can 
occur concurrently with $\chsnu$ or $\chlep$, with branching ratios 
which depend on the degree of mixing, causing changes in the details 
of the limit.

For $\atau = 0$, $\chstau$ decays can occur when $\chsnu$ is kinematically 
forbidden, and the  effects are only noticeable in the limit near the 
corridor. For $\atau = +1~\Tcsq$, the $\Pstaup$ is heavier than 
for $\atau = 0$, and there are few $\chstau$ decays.  A larger difference 
is observed for $\atau = -1~\Tcsq$, as the $\Pstaup$ is light and $\chstau$ 
decays occur with a high branching ratio.  This allows the corridor
to be excluded but weakens the exclusion for lower $\msnu$.

The effects of stau mixing depend on the mixing parameters and field
content of the chargino.  
The most conservative limit is found by using the lowest efficiency.
When the chargino decays to three-body final 
states, this is obtained when the highest possible branching ratio to final 
states with taus occurs, thereby maximising the impact of stau mixing.
This is achieved by varying $M_{\Pstaup}$ and the mixing angle to obtain the 
highest branching ratio for three-body decays to $\tau\nu\Pchi$ (without 
relying on constraints on $\atau$).  This limit is derived for 
$\mu=-500~\Gcsq$, where a branching ratio of $\sim 85\%$ for $\chtau$ for all
sneutrino masses is obtained, for $\msnu > \mcha$.  Since the chargino 
selections have good sensitivity for final states with  taus 
(see Fig.~\ref{fig:eff-cha}), the limit in this ``maximal impact'' case
is not very different from the other examples.  
The result is shown  as the hatched region in Fig.~\ref{fig:stau_3_A}.

\begin{figure}
\begin{center}
\mbox{\epsfig{file=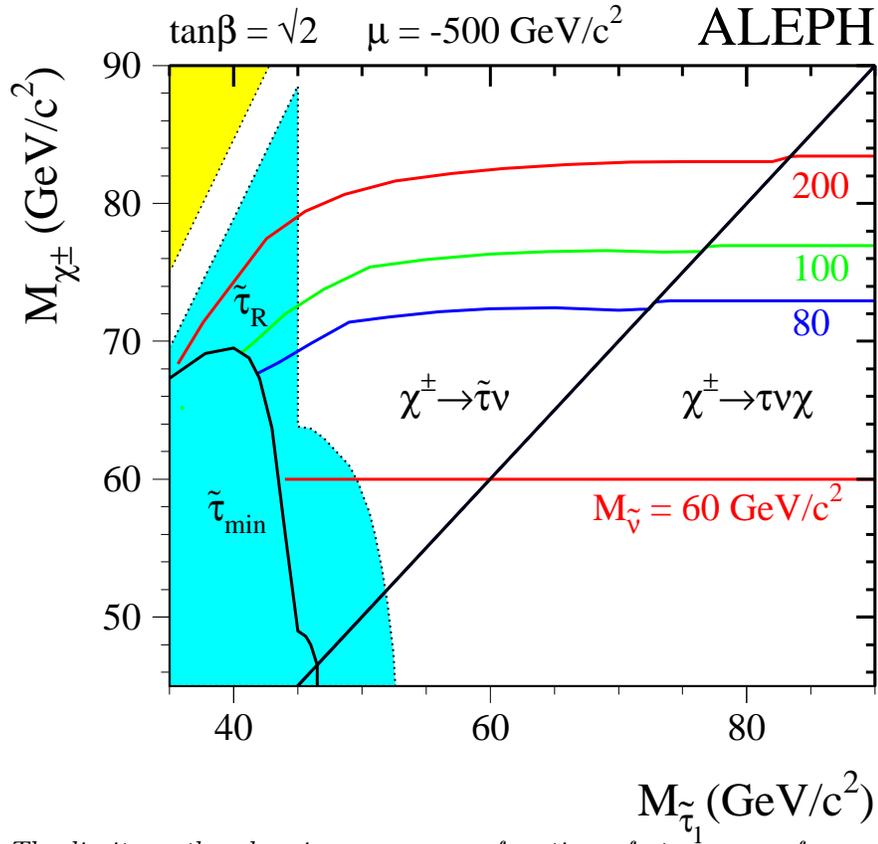,height=15cm%
,bbllx=5mm,bblly=50mm,bburx=175mm,bbury=250mm}}
\caption{\em The limit on the chargino mass as a function of stau mass,
for several values of the sneutrino mass, for
 $\mu=-500~\Gcsq$ and $\tanb = \sqrt{2}$.  
The limits from direct searches
for staus~\protect\cite{sleptons} are also indicated, where the area labelled
$\tilde{\tau}_{R}$ is excluded for pure right-staus, and the area labelled
$\tilde{\tau}_{\mathrm{min}}$ is the most conservative limit.
The light shaded triangular region in the upper left corner corresponds
to $\mchi > M_{\Pstaup}$ and is not considered here.
\label{fig:chastau}}
\end{center}
\end{figure}

The limit on the chargino mass when the requirement of three-body decays is 
released is shown in Fig.~\ref{fig:chastau}, as a function of the $\Pstaup$ 
mass for a series of sneutrino masses.  The transition from three-body 
($\tau\nu\Pchi$) to two-body ($\Pstaup\nu$) final states is marked by the 
diagonal line; the bound falls by a small amount due to the increase of the
final states with taus from $\sim 85\%$ to $100\%$.  Allowing
for $\chstau$ decays degrades the limit by only a few $\Gcsq$ as long
as $M_{\Pstaup} - M_{\Pchi} \ga 10~\Gcsq$.  As before, the mixing angle has
been varied to obtain the lowest bound on the chargino mass. Generally 
this means that the $\Pstaup$ has a high left-stau component, leading 
to $100\%$ $\chstau$ decays.  When $\chsnu$ decays are possible, the most 
conservative scenario is allowing $\chsnu$ decays to dominate, as this opens 
up the corridor discussed previously, and the limit on the chargino mass is 
$\mcha = \msnu$ (as shown in Fig.~\ref{fig:chastau}, for $\msnu = 60~\Gcsq$).
  This corresponds to requiring the $\Pstaup$ to have a 
maximal right-stau component.  Limits from direct searches for staus can be 
invoked in that case, as shown in Fig~\ref{fig:chastau} 
(``$\tilde{\tau}_{\mathrm R}$'').  Also shown is the limit for staus which 
decouple from the Z (``$\tilde{\tau}_{\mathrm{min}}$''), which gives the most 
conservative limit from the stau search.

\subsubsection{Non-universal scalar masses}
\label{non-gut-sleptons}

Interpretation of search results often rely on assumptions according
to a model.  The assumption of a universal slepton mass at the GUT scale
was made in the discussion of stau mixing effects for specific values
of $\atau  (0,\pm 1~\Tcsq)$.  However, in chargino decays, the sneutrinos 
and left sleptons are most relevant; the pure right sleptons do not play 
an important role.  Therefore, theoretical constraints relating the masses 
of the left and right sleptons and sneutrinos can be dropped, retaining 
simply $M_{\tilde{\ell}L}^2 = M_{\tilde{\nu}}^2 - M_{\mathrm W}^{2}\cos2\beta$,
which is guaranteed by gauge invariance.  Equal masses among slepton 
generations are assumed.

In this framework, the results previously derived from the chargino search 
are still valid apart from the stau mixing effects.  The limit in the
``maximal impact'' case is valid without requiring assumptions on a universal
scalar mass, as the most conservative limit is found independently of the
mass of the right-stau.  The requirement of three-body decays is retained for 
the ``maximal impact'' definition, for convenience in generalising the result.
As shown in Fig.~\ref{fig:chastau}, there is little change in the limit when
two-body decays are allowed. 

\subsubsection{Non-universal gaugino masses}

\begin{figure}
\begin{center}
\unitlength=1mm
\begin{picture}(150,10)(0,0)
\put(-3,94){\makebox(0,0){a)}}
\put(83.5,94){\makebox(0,0){b)}}
\end{picture}
\mbox{\epsfig{file=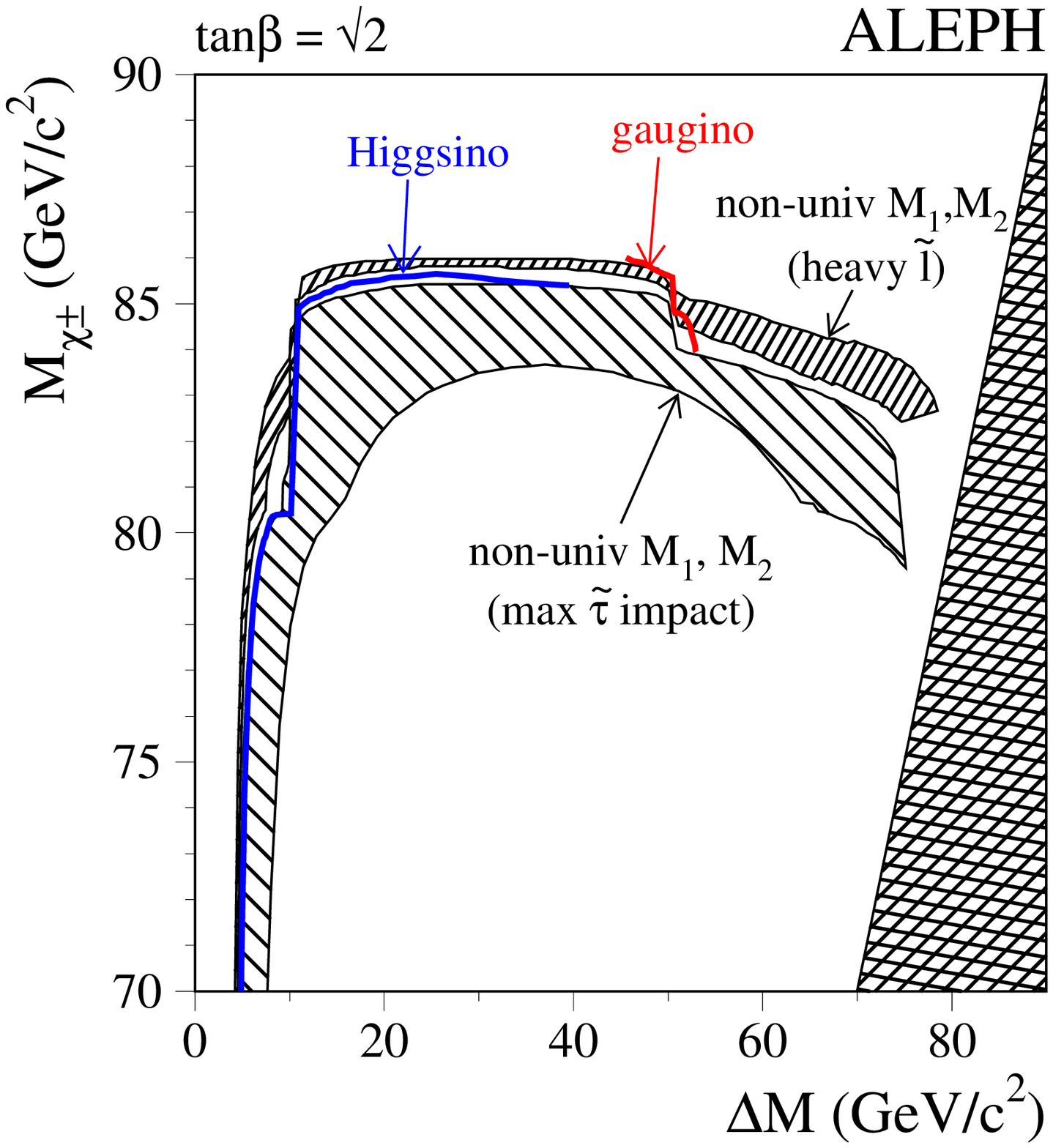,height=9cm%
,bbllx=335mm,bblly=45mm,bburx=335mm,bbury=220mm}}
\mbox{\epsfig{file=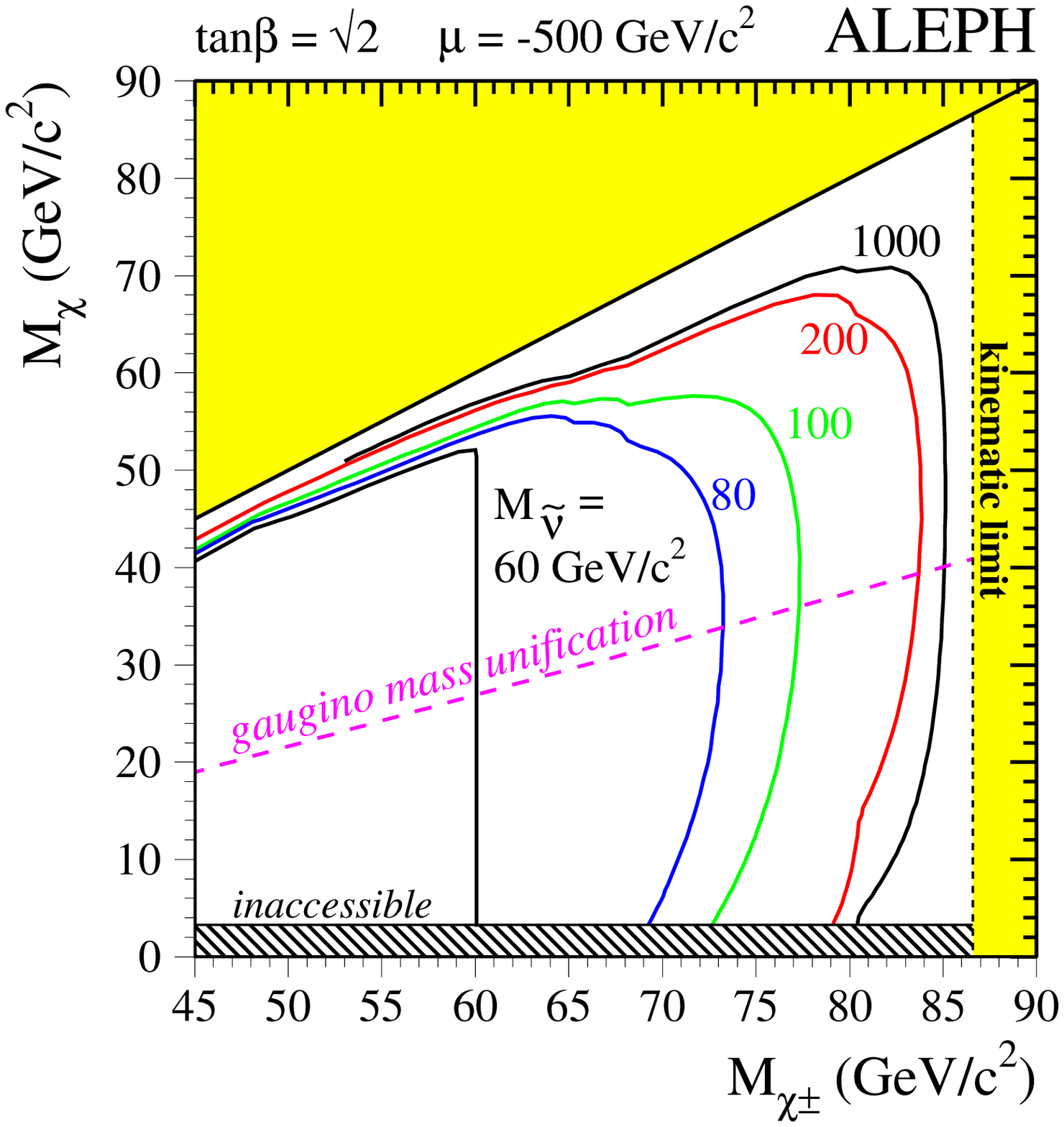,height=9.5cm%
,bbllx=166mm,bblly=44.2mm,bburx=166mm,bbury=219.2mm}}
%
\caption{\em a) The limit on the chargino mass as a 
function of $\dm$, for $\msnu \geq 200~\Gcsq$.
The thick solid curves indicate the limit 
 in the 
Higgsino and gaugino regions, assuming gaugino mass unification. 
The hatched regions reflect the spread in the limits if 
 gaugino mass unification is
relaxed, as $\mu$ is varied from $-80$ to $-500~\Gcsq$,
for heavy 
sleptons, and maximising the impact of stau mixing.
 b) The limit in the ($\mcha,
\mchi$) plane, relaxing gauge unification relations for the gaugino and
slepton masses, for several values of $\msnu$. 
The dashed curve indicates the limit if the gaugino mass unification relation is
assumed.  
The inaccessible region for very low $M_{\Pchi}$ can not be attained 
by relaxing the gauge unification relation.
\label{fig:deltaM}}
\end{center}
\end{figure}

If the unification relation between $M_{1}$ and 
$\Mp$ is assumed, the mass difference between the chargino and 
lightest neutralino 
depends on the parameters of the MSSM: in the gaugino region, $\dm$ is
 $\simeq \mcha/2$; for 
low negative $\mu$ and $\Mp$, $\dm$  can 
be higher; 
in the Higgsino region $\dm$ becomes very small.  
Fig.~\ref{fig:deltaM}a shows the limit on the chargino mass as a function 
of $\dm$ throughout the range of $\dm$ which can be attained 
in the gaugino and 
Higgsino region, for heavy sleptons ($\msnu \geq 200~\Gcsq$). 

If the gauge unification condition is relaxed, the tight correspondence 
between the $\PCha$ and $\Pchi$ masses in the gaugino region can be
broken.
Varying $M_1$ and $\Mp$ independently, the limit on the chargino
mass is displayed as a function of $\dm$ in Fig.~\ref{fig:deltaM}a, for
$\tanb = \sqrt{2}$.  The two hatched bands show the spread in the limit as
$\mu$ is varied between $-500$ and $-80~\Gcsq$, with one band calculated 
assuming heavy sleptons and the other by 
maximising the impact of stau mixing (for three-body decays) 
as defined in Section~\ref{non-gut-sleptons}.  
 The plot shows the
range $0.01 < \alpha < 10$ which is required to attain the entire 
range of $\dm$, where $\alpha$ is defined by
$\alpha = M_{1}/(\frac{5}{3}\tan^{2}\theta_{W}\Mp)$.  
 (This range of $\alpha$ is larger than is expected in typical
SUSY GUT's.) 
The 
reach of the search into the 
region of very high $\dm$ is shown; for a nearly massless neutralino, 
the limit on the chargino mass is $82~\Gcsq$.
These limits change little with $\tanb$, and are valid for high 
$\mzero$ $({\geq} 200~\Gcsq)$.

An excluded region in the  ($\mcha,\mchi$) plane can be derived for a series of
sneutrino masses, as 
shown in Fig.~\ref{fig:deltaM}b.  
In addition to dropping gaugino
mass unification, 
scalar mass unification is relaxed, and the maximum impact of stau
mixing requiring three-body decays is taken into account.

\section{Conclusion}
\label{sec:conclusions}

The data recorded with the ALEPH detector at centre-of-mass energies of 
161, 170, and 172~GeV have been examined for signals of 
chargino and neutralino production.  Selections sensitive to topologies 
arising from chargino
production were developed for a wide range of mass difference
between the chargino and lightest neutralino, and 
especially for very high and very 
low $\dm$.
  Additional selections were developed
for topologies 
arising from chargino and heavier neutralino decays when sleptons are light.
In all of the chargino analyses, 9.5 events were expected,
and in the neutralino analyses, 5.7 events were expected, with some overlap
between the background expectations, giving a total of 13 events expected. 
  A total of 15 candidate events is observed in the
data.
These events are consistent with Standard Model processes, 
giving no evidence of a signal.

Limits at $95\%$ C.L. on the production of charginos and neutralinos 
have been derived, and 
bounds placed on the parameters 
of the MSSM.  The diversity of topological 
selections
and wide range of sensitivity allow interpretation under a variety of model
assumptions.  
Assuming unification of gaugino masses and that sleptons are heavy,
limits are set on the chargino mass, for $\tanb = \sqrt{2}$.
The limit is $85.5~\Gcsq$ in the gaugino region ($\mu = -500~\Gcsq$),  
and 
$85.0~\Gcsq$ in the Higgsino region ($\Mp = 500~\Gcsq$).
The addition of the neutralino
bounds allows the exclusion of charginos beyond the kinematic limit for 
chargino pair production, for moderate $\Mp$ and low $\tanb$.  The combination of
chargino and neutralino searches extends the exclusion in the extreme
Higgsino region, giving a lower limit on the chargino mass of
$79~\Gcsq$ for $\dm \geq 5~\Gcsq$ ($\Mp < 1200~\Gcsq$), 
and $\tanb = \sqrt{2}$.

The search results have been interpreted 
also in the case of light sleptons.
The neutralino search gives powerful exclusion limits in the mixed and gaugino
region. The effect
of stau mixing has been 
investigated in detail,
and limits derived under various
conditions. 
A limit has been obtained with few assumptions about the slepton sector
by maximising the impact of stau mixing.

The gaugino mass unification condition has been relaxed, and limits 
derived for a wide range of $\dm$, corresponding to extreme violations
of the gaugino mass relation.
In addition, scalar mass unification relations have been
 released, 
and limits derived for a range of sneutrino masses, which
are independent of 
assumptions on the mass relations among sleptons and gauginos at the GUT scale,
requiring only that $\mcha < \msnu, M_{\Pstaup}$.

\section*{Acknowledgements}

We thank and congratulate our colleagues in the accelerator divisions for the 
successful startup of LEP2.
We gratefully acknowledge the support of the engineers and technicians in 
our home institutions.
Those of us from non-member states thank CERN for its
hospitality and support.




\newpage


\begin{thebibliography}{9}


\bibitem{SUSYTH} 
H.P.~Nilles, Phys. Rep. {\bf 110} (1984) 1;  \\
H.E.~Haber and G.L.~Kane, Phys. Rep. {\bf 117} (1985) 76; \\
M.~Chen, C.~Dionisi, M.~Martinez and X.~Tata, Phys. Rep. {\bf 159} (1988) 201; \\
R.~Barbieri, Riv. Nuovo Cim. {\bf 11} No. 4 (1988) 1.



\bibitem{sleptons} ALEPH Collaboration,  CERN {\bf PPE/97-056},
to be published in  Phys. Lett. {\bf B}.

\bibitem{stops} ALEPH Collaboration,  CERN {\bf PPE/97-084}, 
to be published in  Phys. Lett. {\bf B}.

\bibitem{Higgs} 
ALEPH Collaboration, CERN {\bf PPE/97-071},
to be published in  Phys. Lett. {\bf B}.


\bibitem{opal-172} 
OPAL Collaboration, CERN {\bf PPE/97-083},
submitted to  Z. Phys. {\bf C}.


\bibitem{RGE}
J.~Ellis and F.~Zwirner, Nucl.~Phys. {\bf B338} (1990) 317;\\
L.~E.~Ibanez and C.~Lopez, {\em ibid.} {\bf B233} (1984) 511;\\
K.~Inoue, A.~Kakuto, H.~Komatsu and S.~Takeshita,
Prog.~Theor.~Phys. {\bf 68} (1982) 927;\\
{\em ibid.} {\bf 71}(1984) 413.


\bibitem{Mtop} CDF Collaboration, Phys. Rev. Lett. {\bf 74} (1995) 2626; \\
D0 Collaboration, Phys. Rev. Lett. {\bf 79} (1997) 1197; \\
D0 Collaboration, FERMILAB-PUB-97/172-E, hep-ex/9706014.

\bibitem{fixed_point} W.A.~Bardeen {\em et al.}, 
Phys. Lett. {\bf B320} (1994) 110.

\bibitem{chaprod} A.~Bartl, H.~Fraas and W.~Majerotto, Z. Phys. {\bf C30}
 (1986) 441; \\
 A.~Bartl, H.~Fraas and W.~Majerotto, Z. Phys. {\bf C41}
 (1988) 475; \\
 A.~Bartl, H.~Fraas, W.~Majerotto and B.~M\"{o}sslacher, Z. Phys. 
{\bf C55} (1992) 257.

\bibitem{chiprod} 
 A.~Bartl, H.~Fraas and W.~Majerotto, Nucl. Phys. {\bf B278} (1986) 1; \\
 S.~Ambrosanio and B.~Mele, Phys. Rev. {\bf D52} (1995) 3900.


\bibitem{SquarkSearch} 
D0 Collaboration, Phys. Rev. Lett. {\bf 75} (1995) 618; \\
CDF Collaboration, Phys. Rev. Lett. {\bf 76} (1996) 2006;\\
D0 Collaboration, Phys. Rev. Lett. {\bf 76} (1996) 2222; \\
CDF Collaboration, Phys. Rev. {\bf D56} (1997) R1357.




\bibitem{nbar}  ALEPH Collaboration, Phys. Lett. {\bf B384} (1996) 427.


\bibitem{AlephDetector} ALEPH Collaboration,
  Nucl. Inst. Meth. {\bf A294} (1990) 121.

\bibitem{AlephPerformances} ALEPH Collaboration,
  Nucl. Inst. Meth. {\bf A360} (1995) 481.

\bibitem{DFGT} C.~Dionisi, K.~Fujii, S.~Giagu and T.~Tsukamoto, in 
{\em Physics at LEP2}, Eds: G.~Altarelli, T.~Sj\"{o}strand, F.~Zwirner,
CERN Report 96--01, Volume 2 (1996) 337.

\bibitem{JETSET} T.~Sj\"{o}strand,
  Comp. Phys. Comm. {\bf 82} (1994) 74.


\bibitem{SUSYGEN} S.~Katsanevas and S.~Melachroinos, in 
{\em Physics at LEP2}, Eds: G.~Altarelli, T.~Sj\"{o}strand, F.~Zwirner,
CERN Report 96--01, Volume 2 (1996) 328.

\bibitem{PHOTOS} E.~Barberio and Z.~Was, Comp. Phys. Comm. {\bf 79}
  (1994) 291.


\bibitem{KORALZ} S.~Jadach and Z.~Was, Comp. Phys. Comm. {\bf 36} (1985) 191.

\bibitem{UNIBAB} H.~Anlauf {\em et al.}, Comp. Phys. Comm. {\bf 79} (1994) 466.

\bibitem{KORALW}  M.~Skrzypek, S.~Jadach, W.~Placzek and Z.~Was,
  Comp. Phys. Comm. {\bf 94} (1996) 216.

\bibitem{PHOT02} J.A.M.~Vermaseren in {\em Proceedings of the IVth 
 international Workshop on Gamma Gamma Interactions},
 Eds G.~Cochard, and P.~Kessler, Springer Verlag, 1980; \\
ALEPH Collaboration, Phys. Lett. {\bf B313} (1993) 509.

\bibitem{PYTHIA} T.~Sj\"{o}strand,
  Comp. Phys. Comm. {\bf 82} (1994) 74;\hfill\break
{\em ibid.}, CERN-TH 7112/93 (1993, revised August 1994).




\bibitem{lepecal} LEP Energy Working Group, LEP Energy Group/97-01;\\
 http://www.cern.ch/LEPECAL/reports/reports.html .

\bibitem{Cousins} 
R.D. Cousins and W.L. Highland, Nucl. Inst. Meth. {\bf A320} (1992) 331-335.

\bibitem{pdg} R.M.~Barnett {\em et al.},
(Particle Data Group), Phys. Rev. {\bf D54} (1996) 1.



\bibitem{bib:paper133} 
 ALEPH Collaboration, Phys. Lett. {\bf B373} (1996) 246.


\bibitem{bib:lsp-133}
 ALEPH Collaboration, Z. Phys. {\bf C72} (1996) 549-559.



\bibitem{opal-lep1} 
 OPAL Collaboration, Phys. Lett. {\bf B377} (1996) 273.


\bibitem{interpretations}
ALEPH Collaboration, 
 ``Update of the Mass Limit for the Lightest Neutralino'',
contribution to the International Europhysics Conference on High Energy
Physics, Jerusalem, Israel, $19 - 26$ Aug. 1997, Ref. no. 594.





\end{thebibliography}
\end{document}